\documentclass[smallextended,numbook,envcountsect,envcountsame,natbib,final]{svjour3}

\smartqed

\usepackage[latin1]{inputenc}
\usepackage[english]{babel}
\usepackage{hyphenat}
\usepackage{amsmath}
\usepackage{amsfonts}
\usepackage{mathrsfs}
\usepackage{colonequals}
\usepackage{amssymb}
\usepackage{amscd}
\usepackage{enumitem}
\usepackage{amsbsy}
\usepackage{setspace}
\usepackage{calc}
\usepackage{makeidx}
\usepackage{url}
\usepackage[refpage]{nomencl}
\usepackage{marvosym}

\spdefaulttheorem{assumption}{Assumption}{\bf}{\it}
\spnewtheorem*{Obs*}{Remark}{\it}{\normalfont}

\newcommand{\bbOne}{1\hspace*{-0.8ex}1}

\journalname{Name of Journal}

\begin{document}

\title{Optimal Portfolio Choice for a Behavioural Investor in Continuous-Time 
Markets}

\author{Mikl\'os R\'asonyi \and Andrea M. Rodrigues}


\institute{M. R\'asonyi (\Letter), A. M. Rodrigues \at
School of Mathematics, University of Edinburgh, Edinburgh EH9 3JZ, Scotland\\
              \email{Miklos.Rasonyi@ed.ac.uk}
}

\date{Received: \today\ / Accepted: \today}
\maketitle

\begin{abstract}
	The aim of this work consists in the study of the optimal investment strategy for a behavioural investor, whose preference towards risk is described by both a probability distortion and an \emph{S}-shaped utility function. Within a continuous-time financial market framework and assuming that asset prices are modelled by semimartingales, we derive sufficient and necessary conditions for the well-posedness of the optimisation problem in the case of piecewise-power probability distortion and utility functions. Finally, under straightforwardly verifiable conditions, we further demonstrate the existence of an optimal strategy.
	


\end{abstract}

\section{Introduction}\label{Introduction}

Portfolio optimisation in a securities market is a classical problem in financial mathematics. Although Expected Utility Theory (or EUT), in which the economic agents' individual preferences towards risk are characterised by a utility function, has been a predominantly used model for decision making under uncertainty since it was introduced by \citet{neumann53}, its basic tenets have been contradicted by empirical evidence. First, instead of thinking about final asset states, as assumed in EUT, investors evaluate wealth in terms of gains and losses. Second, according to the expected utility hypothesis, the investors' risk attitude is constant (generally globally risk averse), whereas in practice people were found to be risk averse when gaining and risk seeking on losses, as well as loss averse (that is, they are considerably more sensitive to losses than to gains). It also turned out that economic agents are not evaluating probabilities objectively, rather they are subjective and tend to overweight small 
probabilities \citep[which may explain the attractiveness of insurance and gambling, as remarked for instance in][]{kahneman79}.

The attempts to incorporate the above psychological findings into economics theory have given rise to the elaboration of alternative approaches regarding investor decision under uncertainty. In particular, \citet{kahneman79} suggested Prospect Theory, which was subsequently developed into Cumulative Prospect Theory, or CPT as it is commonly known \citep{tversky92}. Within this framework, the economic agents are no longer assumed to analyse the information and make decisions in the face of uncertainty in a fully rational manner, and the main axioms of this theory consist of the existence, for each investor: of a reference point (according to which gains and losses are defined); of a prospect value function (corresponding to the utility function of EUT) that is concave for gains and convex for losses (and steeper for losses than for gains to reflect loss aversion); and finally of a probability weighting function which distorts the probability measure in a nonlinear way.

In this work, we shall therefore study the optimal portfolio problem for an agent behaving consistently with CPT. This proves, however, to involve a considerable degree of complexity, especially since many of the mathematical tools used in EUT are no longer applicable in CPT. In particular, we are dealing with an overall \emph{S}-shaped utility function, so standard convex optimisation techniques are impossible to use. Moreover, due to the fact that probabilities are distorted, Choquet integrals now appear in the definition of the payoff functional to be maximised, thus the dynamic programming principle cannot be used either, as remarked in \citet{jin2008} and \citet{carassus2011}, because dynamic consistency does not hold \citep[in fact, there is not even an uncontroversial definition of the conditional Choquet expectation; we refer for example to][]{kast08}. We note further that it may be the case that the optimisation problem is ill-posed (that is, the supremum over admissible strategies may not be finite), even in seemingly innocuous cases, which brings to light additional issues.

This question of ill-posedness is actually a frequent one in the context of CPT, and is discussed in a discrete-time setting by \citet{he2011} for a single-period model, and also in \citet{carassus2011} in multiperiod (and generically incomplete) models. Furthermore, the problem of well-posedness assuming a complete continuous-time market and It\^{o} processes for asset prices arises in \citet{jin2008}. Positive results in \citet{jin2008}, however, are provided only under conditions that are not easily verifiable and their economic interpretation is not obvious. 

In our work we study the issue of well-posedness in similar models. We focus essentially on the case where the utilities and probability distortions are given by piecewise-power functions, such as the ones suggested in \citet{tversky92}. Some related investigations
have been carried out in \cite{cd11}, but they use the risk-neutral (instead of the physical) probability in the definition of the objective function, which leads to a problem that is entirely different from ours. The results of \citet{jin2008} do not seem to cover
the case of power-like distortions \citep[see the discussion in subsection 6.2 of][]{jin2008}.

Additionally, once it is ensured that the optimisation problem with which we are dealing is a well-posed one, it still remains to show that an optimal solution does in fact exist. Therefore, the issue of attainability, which is an interesting and delicate problem on its own, is also addressed in this work. The existence of an optimal portfolio for a CPT investor is established in \citet{bernard2010} and in \citet{carassus2011}, respectively in a one-period economy with one risk-free asset and one risky asset, and in a multiperiod market. The optimal investment strategy for an investor exhibiting loss aversion in a complete continuous-time market is obtained by \citet{berkelaar04}, but no probability distortion is considered in the paper, which makes the problem considerably simpler. See also \cite{reichlin2011}. In \citet{jin2008} and \citet{carlier-dana2011} the optimal terminal wealth positions are explicitly derived in some specific cases; however, we again note that the conditions provided are not easy 
to verify and the probability distortions which are considered are fairly specific ones. All the three papers \citet{berkelaar04}, \citet{jin2008} and \citet{carlier-dana2011} assume that the attainable wealth  of an admissible portfolio is bounded from below, which looks rather unnatural when we recall that, in EUT, optimal strategies typically lead to arbitrarily large losses. In this paper we present existence results for the optimiser in a more natural class of portfolio strategies.

A brief summary of the present paper is as follows. Section~\ref{ch:Model} is dedicated to a brief recollection of definitions and results, as well as to the introduction of the general assumptions in force throughout the paper. We start by specifying the financial market model, which is followed by the mathematical formulation of the main elements of cumulative prospect theory. We then define the CPT objective function and lastly the optimal portfolio choice problem for a behavioural investor in a continuous-time economy is stated. Next, in Section~\ref{ch:WellPosed} we focus on the issue of well-posedness, and examine sufficient and necessary conditions for the problem to be well-posed, while in Section~\ref{ch:Existence} the existence of an optimal trading strategy under suitable conditions is investigated. An example is provided in the following section to demonstrate that the obtained results hold in important model classes. In the last section, which concludes the paper, we summarise the main results 
and contributions of this work. We also include some suggestions on further research to improve these results. Finally, to help readability, most of the proofs are collected in the appendix.


\section{Model Description}\label{ch:Model}

\subsection{The Market}\label{TheMarket}

Let the current time be denoted by $0$ and let us fix a nonrandom planning horizon, or maturity, $T \in \left(0,+\infty\right)$. We are thus considering a continuous\hyp{}time economy with trading interval $\left[ 0,T \right]$. In addition, we are assuming the financial market to be liquid and frictionless, that is, all costs and restraints associated with transactions are non\hyp{}existent, investors are allowed to short\hyp{}sell stocks and to borrow money, and it is always possible to buy or sell an unlimited number of shares of any asset. As usual, the uncertainty in the economy is characterised by a complete probability space $\left( \Omega,\mathscr{F},\mathbb{P} \right)$, where $\mathscr{F}$ is a $\sigma$\hyp{}algebra on the sample space $\Omega$, and $\mathbb{P}$ is the underlying probability measure (to be interpreted as the physical probability) on $\left(\Omega,\mathscr{F}\right)$. Moreover, all the information accruing to the agents in the economy is described by a filtration $\mathbb{F}=\left\{\
mathscr{F}_{t}\text{; } 0 \leq t \leq T\right\}$ satisfying the \emph{usual conditions} of right-continuity and saturatedness. %
Finally, we assume for simplicity that the $\sigma$-algebra $\mathscr{F}_{0}$ is $\mathbb{P}$-trivial 
and also that $\mathscr{F}=\mathscr{F}_{T}$.

The financial market consists of $d+1$ continuously traded assets, where $d$ is a positive integer. We shall assume that there exists a risk-free asset, called 
\emph{money market account}, %
with price $S^{0}_{t}$ per share at time $t \in \left[0,T\right]$. 
Also, we consider $d$ risky securities, which we refer to as \emph{stocks}\index{stock},\footnote{Note that it does not necessarily have to be a common stock, it can also be referring to a commodity, a foreign currency, an exchange rate or a market index.} %
whose price evolution is modelled by a nonnegative, $\mathbb{R}^{d}$-valued (not necessarily continuous) semimartingale $S=\left\{S_{t}=\left(S^{1}_{t},\ldots,S^{d}_{t}\right), \mathbb{F}\text{; } 0 \leq t \leq T\right\}$, where $S^{i}_{t}$ represents the price of asset $i\in \left\{1,\ldots,d\right\}$ at time $t$ expressed in units of the riskless asset. 
In order to keep the notation simple, we find it convenient to suppose, without loss of generality, that the money market is constant, $S^{0}_{t}\equiv 1$ (economically speaking, this means that the \emph{interest rate} is zero). Hence, we work directly with discounted prices and we say that the market is \emph{normalised}.

Assume further that the set of \emph{equivalent martingale measures} (e.m.m. for short) for $S$ is not empty, i.e., there exists at least one probability measure $\mathbb{Q}$ on $\left(\Omega,\mathscr{F}\right)$ such that $\mathbb{Q}$ is equivalent to $\mathbb{P}$ (we write $\mathbb{Q} \sim \mathbb{P}$), and such that the (discounted) price process $S$ is a local martingale with respect to $\mathbb{Q}$. Let the random variable (r.v.) $\rho$ designate the Radon-Nikodym derivative of $\mathbb{Q}$ with respect to $\mathbb{P}$, $\rho \colonequals \frac{d\mathbb{Q}}{d\mathbb{P}}$. We shall impose the following technical assumptions throughout.
\begin{assumption}\label{as:CDFW}
	The cumulative distribution function (CDF) of $\rho$ under $\mathbb{Q}$,\footnote{We recall that, for every real number $x$, the cumulative distribution function of $\rho$ with respect to the probability measure $\mathbb{Q}$ is given by $F_{\rho}^{\mathbb{Q}}\!\left(x\right)=\mathbb{Q}\!\left(\rho \leq x\right)$.} which we denote by $F_{\rho}^{\mathbb{Q}}$, is continuous. Moreover, both $\rho$ and $1/\rho$ belong to $\mathscr{W}$, where $\mathscr{W}$ is defined as the family of all real-valued and $\mathscr{F}_{T}$-measurable random variables $Y$ satisfying $\mathbb{E}_{\mathbb{P}}\!\left[\left|Y\right|^{p}\right]<+\infty$ for all $p>0$.
\end{assumption}

\begin{assumption}\label{as:Ustar}
	There exists an $\mathscr{F}_{T}$-measurable random variable $U_{*}$ such that, under $\mathbb{P}$, $U_{*}$ follows a uniform distribution on the interval $\left(0,1\right)$ and is independent of $\rho$.
\end{assumption}
For a thorough discussion on the introduction of an external source of randomness, we refer to \citet[][Section~5]{carassus2011}, and we shall see below that this assumption holds in important model classes. Finally, we also fix an integrable scalar-valued $\mathscr{F}_{T}$-measurable random variable $B$ as our reference point.

Now let $\Phi$ denote the class of all \emph{portfolios} or \emph{trading strategies} over the time interval $\left[0,T\right]$ that are \emph{self\hyp{}financed}. These consist of all the $\left(d+1\right)$\hyp{}dimensional predictable stochastic processes $\phi=\left\{\phi_{t}=\left(\phi^{0}_{t},\phi^{1}_{t},\ldots,\phi^{d}_{t}\right)\text{; } 0\leq t \leq T\right\}$, where $\phi^{i}_{t}$ represents the amount of shares of the $i$\hyp{}th asset possessed at time $t$, which are $S$-integrable (that is, for which the stochastic integral $\int_{0}^{T} \phi_{t} dS_{t}$ is well\hyp{}defined and then again a semimartingale%
) and whose associated \emph{wealth process} $\Pi^{\phi}=\left\{\Pi^{\phi}_{t}\text{; } 0\leq t\leq T\right\}$, given by $\Pi^{\phi}_{t} \colonequals \sum_{i=0}^{d} \phi^{i}_{t} S^{i}_{t}$, $0\leq t \leq T$, satisfies the self\hyp{}financing condition
\begin{equation}\label{eq:selffin}
	\Pi^{\phi}_{t}=\Pi^{\phi}_{0}+\int_{0}^{t} \phi_{s} dS_{s},
\end{equation}
for every $0 \leq t \leq T$. %
To rule out pathologies, such as doubling schemes or suicide strategies \citep[see, e.g.,][]{harrison1981}, we must restrict our attention to a subset $\Psi \subseteq \Phi$ of admissible strategies. A condition of admissibility which is commonly imposed in the literature \citep[and adopted in][]{berkelaar04,jin2008,carlier-dana2011} is that the (discounted) wealth process corresponding to the portfolio should be bounded uniformly from below by a real constant \citep[such a strategy is also known as a \emph{tame} strategy, see for instance][]{karatzas98}. Although this restriction, which reflects the existence of a credit limit, is not an unrealistic one, it looks rather unnatural \citep[especially when we recall that, in general in EUT, we cannot hope to find an optimal strategy that leads to bounded losses; we refer for instance to][]{schachermayer01}. Therefore, in this work, we choose to adopt the following.
\begin{definition}\label{def:admissible}
	A self-financed trading strategy is said to be \emph{admissible} if its (discounted) wealth is a martingale under $\mathbb{Q}$.
\end{definition}

Finally, we recall that a \emph{European contingent claim} settling at time $T$ is represented by an $\mathscr{F}_{T}$-measurable random variable $H$, and we say that the claim is \emph{hedgeable} if there exists a portfolio $\phi \in \Psi$ such that its associated wealth process satisfies $\Pi^{\phi}_{T}=H$ (we call such $\phi$ a \emph{replicating} strategy). Hereafter, we shall assume the following, which will be needed frequently.
\begin{assumption}\label{as:kindcompl}
	The integrable random variable $B$ and all $\sigma\!\left(\rho,U_{*}\right)$-meas\-ur\-a\-ble random variables in $L^{1}\!\left(\mathbb{Q}\right)$ (i.e., integrable with respect to the measure $\mathbb{Q}$) are hedgeable.
\end{assumption}
This additional assumption is a kind of completeness hypothesis on the market, although for a certain type of claims only, thus being weaker than the usual notion of market completeness. Indeed, it is trivial to see that this assumption holds within a complete market framework, but it can also be satisfied for incomplete financial models, as shown by the example provided in the upcoming Section~\ref{ch:Example}.

\subsection{The Investor}

As described in the Introduction, we analyse a representative economic agent, with a given initial capital $x_{0} \in \mathbb{R}$, who behaves in accordance with (Cumulative) Prospect Theory, introduced and developed by \citet{kahneman79,tversky92}. We are dealing with a small investor, whose behaviour has no effect on the movement of asset prices.

Firstly, according to this framework, the investor is assumed to have reference point $B$, with respect to which payoffs at the terminal time $T$ are evaluated.

Secondly, rather than a (concave) utility function, agents now use an \emph{S}\hyp{}shaped utility function $u:\mathbb{R} \rightarrow \mathbb{R}$ \citep[or, to follow the nomenclature of][prospect value function]{tversky92}, concave for gains and convex for losses, in order to express their risk\hyp{}aversion in gains and their risk\hyp{}seeking attitude with respect to losses. Mathematically speaking, this means that we are considering two continuous, strictly increasing and concave utility functions, $u_{+},u_{-}:\left[\left.0,+\infty\right)\right. \rightarrow \left[\left.0,+\infty\right)\right.$, satisfying $u_{+}\!\left(0\right)=u_{-}\!\left(0\right)=0$, and we take $u\!\left(x\right) \colonequals u_{+}\!\left(\left[x-B\right]^{+}\right)\bbOne_{\left[\left.B,+\infty\right)\right.}\!\left(x\right)-u_{-}\!\left(\left[x-B\right]^{-}\right)\bbOne_{\left(-\infty,B\right)}\!\left(x\right)$.\footnote{Note that here $x^{+}=\max\left\{x,0\right\}$ and $x^{-}=-\min\left\{x,0\right\}$.}

Lastly, the CPT investor has a distorted perception of the actual probabilities, which is represented by the probability weighting functions $w_{+}$ and $w_{-}$, both mapping from $\left[0,1\right]$ to $\left[0,1\right]$. These are assumed to be continuous and strictly increasing, with $w_{+}\!\left(0\right)=w_{-}\!\left(0\right)=0$ and $w_{+}\!\left(1\right)=w_{-}\!\left(1\right)=1$.

\subsection{The Optimal Portfolio Problem}

Given an arbitrary nonnegative random variable X, let us start by defining the values $V_{+}\!\left(X\right)$ and $V_{-}\!\left(X\right)$, given respectively by the \emph{Choquet integrals}\index{Choquet integral} of $X$ with respect to the \emph{capacities}\index{capacity} $w_{+} \circ \mathbb{P}$ and $w_{-} \circ \mathbb{P}$ (these are non necessarily additive measures, which can be regarded as the agent's subjective measures of the likelihood of gains and losses), as follows
\begin{eqnarray}
	V_{+}\!\left(X\right) &\colonequals& \int_{0}^{+\infty} w_{+}\!\left(\mathbb{P}\!\left\{u_{+}\!\left(X\right)>y\right\}\right) dy,\label{eq:Vp}\\
	V_{-}\!\left(X\right) &\colonequals& \int_{0}^{+\infty} w_{-}\!\left(\mathbb{P}\!\left\{u_{-}\!\left(X\right)>y\right\}\right) dy.\label{eq:Vm}
\end{eqnarray}
It can be easily verified that $V_{\pm}\!\left(c\right)=u_{\pm}\!\left(c\right)$ for any constant $c \in \left[\left.0,+\infty\right)\right.$. Furthermore, $V_{+}\!\left(X\right)\leq V_{+}\!\left(Y\right)$ and $V_{-}\!\left(X\right)\leq V_{-}\!\left(Y\right)$ for any random variable $Y$ satisfying $0 \leq X\leq Y$ almost surely (a.s.\ for short). Finally, taking now $X$ to be any random variable (not necessarily nonnegative), whenever $V_{-}\!\left(X^{-}\right)<+\infty$ we set
\begin{equation}\label{eq:V}
	V\!\left(X\right) \colonequals V_{+}\!\left(X^{+}\right)-V_{-}\!\left(X^{-}\right),
\end{equation}
which is also non\hyp{}decreasing in the sense described above.

The continuous\hyp{}time portfolio choice problem for an investor with CPT preferences then consists of selecting the optimal trading strategies, from the set $\mathscr{A}\!\left(x_{0}\right)$ of all \emph{feasible controls} to be defined later, in terms of maximising the expected distorted payoff functional $V\!\left(\Pi^{\phi}_{T}-B\right)$, which can be mathematically formalised as follows:
\begin{equation}\label{eq:optport}
	\textrm{Maximise } \left\{V\!\left(\Pi^{\phi}_{T}-B\right)=V_{+}\!\left(\left[\Pi^{\phi}_{T}-B\right]^{+}\right)-V_{-}\!\left(\left[\Pi^{\phi}_{T}-B\right]^{-}\right)\right\}
\end{equation}
over $\phi \in \mathscr{A}\!\left(x_{0}\right)$.


We conclude this chapter by noticing that, in general, when studying optimisation problems such as~\eqref{eq:optport} above, some issues may arise, namely those of \emph{well\hyp{}posedness} and \emph{attainability}. Addressing them shall be the purpose of the subsequent sections.


\section{Well-Posedness}\label{ch:WellPosed}

We begin this section by introducing the following.
\begin{definition}\label{wellposed}
	A maximisation problem is termed \emph{well-posed}\index{well-posedness} if its supremum is finite. Problems that are not well-posed are said to be \emph{ill-posed}.
\end{definition}

For an ill-posed problem, maximisation does not make sense. Intuitively speaking, it means that the investor can obtain an arbitrarily high degree of satisfaction from the trading strategies that are available in the market. For this reason, before searching for the optimal portfolio for the behavioural investor of Section~\ref{ch:Model}, we first need to identify and exclude the ill-posed cases. The intent of this section is to provide a detailed study of the well-posedness of the portfolio selection problem~\eqref{eq:optport}, and to find conditions ensuring well-posedness.

\subsection{Feasible Portfolios}\label{FeasiblePort}

In order to be able to tackle the problem's well-posedness, we must start by specifying the family of all feasible trading strategies, $\mathscr{A}\!\left(x_{0}\right)$. Because the investor is assumed to have an initial wealth of $x_{0}$, we restrict ourselves to the set of trading strategies $\phi$ in $\Psi$ for which $\Pi^{\phi}_{0}=x_{0}$. We note that, since the process $\Pi^{\phi}$ is a $\mathbb{Q}$-martingale, in particular this implies that $\mathbb{E}_{\mathbb{P}}\!\left[\rho \Pi^{\phi}_{T}\right]=\mathbb{E}_{\mathbb{Q}}\!\left[\Pi^{\phi}_{T}\right]=x_{0}$. Additionally, we have to impose the condition that the quantity $V_{-}\!\left(\left[\Pi^{\phi}_{T}-B\right]^{-}\right)$ is finite, to ensure that the functional $V$ given by~(\ref{eq:V}) is well-defined.

Given that, by virtue of Assumption~\ref{as:kindcompl}, $B$ admits a replicating portfolio, it is easy to see that the optimal portfolio problem can be reduced to one with reference level equal to zero \citep[see][Remark~2.1]{jin2008}. Hence, for simplicity and without loss of generality, from now on we set $B=0$.

We further remark that if there exists an hedgeable claim $H$ so that $V_{+}\!\left(H^{+}\right)=+\infty$ \citep[for an example that such a claim can actually be found, we refer to][Example~3.1]{jin2008}, then the problem is clearly ill-posed.

Also, we would like to ensure that the set of admissible portfolios is not empty, that is, there exists at least one portfolio satisfying all the conditions imposed above. This is guaranteed by the following, obvious result.
\begin{lemma}\label{notempty}
	Consider the claim $X_{0}=\frac{x_{0}}{\rho}$. Then $X_{0}$ is hedgeable and a replicating portfolio for this claim is feasible for problem~\eqref{eq:optport}.\qed
\end{lemma}

We end this short discussion by conventioning that, whenever $X$ is a claim admitting a replicating portfolio that belongs to the set $\mathscr{A}\!\left(x_{0}\right)$, by abuse of language we may write ``$X$ is in $\mathscr{A}\!\left(x_{0}\right)$'' or ``$X$ is feasible for~\eqref{eq:optport}''.


\subsection{Piecewise-Power Utilities and Probability Distortions}\label{piecewisepower}

Motivated by the work of \citet{carassus2011}, throughout this subsection, and for the rest of the paper, we shall assume that the utility functions and the probability distortions are power functions\index{piecewise-power} of the form
\begin{eqnarray}
	u_{+}\!\left(x\right)\colonequals x^{\alpha}\quad \textrm{and}\quad u_{-}\!\left(x\right)\colonequals x^{\beta}, & & \textrm{for all } x \in \left[\left.0,+\infty\right)\right.,\label{eq:poweru}\\
	w_{+}\!\left(x\right)\colonequals x^{\gamma}\quad \textrm{and}\quad w_{-}\!\left(x\right)\colonequals x^{\delta}, & & \textrm{for all } x \in \left[0,1\right],\label{eq:powerw}
\end{eqnarray}
where the constraints $0<\alpha,\,\beta, \gamma, \delta \leq 1$ are imposed on the parameters.

\begin{remark}\label{constraints}
	By requiring that $0<\alpha,\,\beta\leq 1$, we ensure that both $u_{+}$ and $u_{-}$ are strictly increasing, as well as concave. Also, we have that the weighting functions $w_{+}$ and $w_{-}$ are strictly increasing if and only if $\delta,\,\gamma>0$. Moreover, by assuming $\gamma, \delta \leq 1$, we get that the probability distortions are also concave and that $w_{\pm}\!\left(x\right)\geq x$ holds for all $x \in \left[0,1\right]$, although more significantly when $x$ is ``close'' to zero (hence capturing the fact that a behavioural agent overweights small probabilities).
	
	Our results apply with trivial modifications to the utility and distortion functions proposed by \citet{tversky92} as well:
	\begin{eqnarray*}
		u_{+}\!\left(x\right)\colonequals c_{1} x^{\alpha} &\textrm{and}& u_{-}\!\left(x\right)\colonequals c_{2} x^{\beta},\\
	w_{+}\!\left(x\right)\colonequals \frac{x^{\gamma}}{\left(x^{\gamma}+\left(1-x\right)^{\gamma}\right)^{1/\gamma}} &\textrm{and}& w_{-}\!\left(x\right)\colonequals \frac{x^{\delta}}{(x^{\delta}+\left(1-x\right)^{\delta})^{1/\delta}},
\end{eqnarray*}
for some $c_{1},c_{2}>0$.
\end{remark}

As stated above, we are concerned with seeking conditions on the parameters under which the portfolio problem is a well-posed one. Once again inspired by \citet{carassus2011}, we start by proving that, as in the incomplete discrete-time multiperiod case, we need to assume $\alpha < \beta$ in order to obtain a well-posed optimisation problem.
\begin{proposition}\label{a>b}
	If $\alpha >\beta$, then the problem~(\ref{eq:optport}) is ill-posed.
\end{proposition}
\begin{proof}
	Suppose that $\alpha>\beta$ and let $U$ be the random variable given by $U\colonequals F_{\rho}^{\mathbb{Q}}\!\left(\rho\right)$. Recalling the assumption that $F_{\rho}^{\mathbb{Q}}$ is continuous, it is then a well-known result that $U$ has uniform distribution on $\left(0,1\right)$ under $\mathbb{Q}$, $U \stackrel{_{\mathbb{Q}}}{\sim} \mathscr{U}\!\left(0,1\right)$.
	
	For each $n \in \mathbb{N}$, we define the nonnegative random variable $Y_{n}\colonequals n\,\bbOne_{A}$, with $A\colonequals \left\{\omega \in \Omega: U\!\left(\omega\right)\geq \frac{1}{2}\right\}$. Then $\mathbb{E}_{\mathbb{Q}}\!\left[Y_{n}\right]=n\,\mathbb{Q}\!\left(A\right)=\frac{n}{2}$ and
	\begin{equation*}
		V_{+}\!\left(Y_{n}\right)=\int_{0}^{+\infty} \mathbb{P}\!\left\{Y_{n}^{\alpha}>y\right\}^{\gamma} dy=\int_{0}^{n^{\alpha}} \mathbb{P}\!\left\{Y_{n}^{\alpha}>y\right\}^{\gamma} dy=n^{\alpha}\,\mathbb{P}\!\left(A\right)^{\gamma}.
	\end{equation*}
	
	Now set $Z_{n}\colonequals \left(n-2 x_{0}\right)\bbOne_{A^{c}}$, for every $n \in \mathbb{N}$.\footnote{We denote by $A^{c}$\nomenclature[A]{$A^{c}$}{Complement of $A$ in $\Omega$} the complement of $A$ in $\Omega$.} Clearly we have that $\mathbb{E}_{\mathbb{Q}}\!\left[Z_{n}\right]=\left(n-2 x_{0}\right) \mathbb{Q}\!\left(A^{c}\right)=\frac{n}{2}-x_{0}$, %
	so $\mathbb{E}_{\mathbb{Q}}\!\left[Y_{n}\right]-\mathbb{E}_{\mathbb{Q}}\!\left[Z_{n}\right]=x_{0}$. Furthermore, computations similar to those above show that
	\begin{equation*}
		V_{-}\!\left(Z_{n}^{+}\right)=\int_{0}^{+\infty} \mathbb{P}\!\left\{\left(Z_{n}^{+}\right)^{\beta}>y\right\}^{\delta} dy=\left(\left[n-2 x_{0}\right]^{+}\right)^{\beta} \mathbb{P}\!\left(A^{c}\right)^{\delta}.
	\end{equation*}
	
	Finally, since $2 x_{0} \leq n_{0}$ for some $n_{0} \in \mathbb{N}$, let us define for each $n \in \mathbb{N}$ the random variable $X_{n}\colonequals Y_{n_{0}+n}-Z_{n_{0}+n}$, which is clearly $\sigma\!\left(\rho\right)$-measurable and bounded from below by $2 x_{0} -n_{0}-n$. It can also be easily checked that $X_{n}^{+}=Y_{n_{0}+n}$ and $X_{n}^{-}=Z_{n_{0}+n}$, so $\mathbb{E}_{\mathbb{P}}\!\left[\rho X_{n}\right]=x_{0}$. Therefore, for each $n$ the r.v. $X_{n}$ is feasible for~\eqref{eq:optport}, however the sequence
	\begin{equation*}
		V\!\left(X_{n}\right)=\left(n_{0}+n\right)^{\alpha} \mathbb{P}\!\left(A\right)^{\gamma}-\left(n_{0}+n-2 x_{0}\right)^{\beta} \mathbb{P}\!\left(A^{c}\right)^{\delta}
	\end{equation*}
	goes to infinity as $n \rightarrow +\infty$ (we recall that $\mathbb{P}\!\left(A\right)>0$ because $\mathbb{P}$ and $\mathbb{Q}$ are equivalent measures), hence $\sup_{\phi \in \mathscr{A}\!\left(x_{0}\right)} V\!\left(\Pi_{\phi}\!\left(T\right)\right)=+\infty$.\qed
	
\end{proof}


\begin{Obs*}
	Suppose that there exists an event $A$, with $\mathbb{Q}\!\left(A\right)=1/2$, for which $\mathbb{P}\!\left(A\right)^{\gamma}>\left[1-\mathbb{P}\!\left(A\right)\right]^{\delta}$ also holds true. Then even in the case where $\alpha=\beta$, 
\begin{equation*}
		V\!\left(X_{n}\right)=n^{\alpha}\left[\left(1+\frac{n_{0}}{n}\right)^{\alpha} \mathbb{P}\!\left(A\right)^{\gamma}-\left(1+\frac{n_{0}-2 x_{0}}{n}\right)^{\alpha} \mathbb{P}\!\left(A^{c}\right)^{\delta}\right]\xrightarrow[n\rightarrow +\infty]{} +\infty,
	\end{equation*}
shows us that the optimisation problem~\eqref{eq:optport} is ill-posed.
	
	We shall now provide an example of a financial market model in which such an event can be found. First, let us define the function $f\!\left(p\right)\triangleq p^{\gamma}-\left(1-p\right)^{\delta}$ for $p\in\left[0,1\right]$. Clearly, there exists some $\epsilon>0$ such that $f\!\left(x\right)>0$ for all $x\in\left(\left.1-\epsilon,1\right]\right.$. On the other hand, choosing $\mu$ to be sufficiently large, we have that
	\begin{equation*}\label{GaussCDF}
		\int_{-\infty}^{-\mu} \frac{1}{\sqrt{2\pi}}e^{-x^{2}/2}\,dx<\epsilon.
	\end{equation*}
	
	Next, let the continuous process $W=\left\{W_{t}\textrm{; } 0\leq t \leq 1\right\}$ be one-dimensional \emph{Wiener process} (with respect to the probability measure $\mathbb{P}$) initialised at zero a.s.. We are assuming the flow of information in the market to be represented by the \emph{natural filtration} of $W$ (which is the augmentation, by all $\mathbb{P}$\hyp{}null sets, of the filtration generated by $W$), $\mathbb{F}^{W}=\left\{\mathscr{F}_{t}^{W}\textrm{; }0\leq t \leq 1\right\}$. We further recall that the interest rate is here assumed to be null. Finally, the dynamics of the price process $S=\left\{S_{t}\text{; } 0 \leq t \leq 1\right\}$ of the risky asset is described, under the measure $\mathbb{P}$, by the Itô process with stochastic differential
	\begin{equation*}\label{eq:stockpricealphaeqbeta}
		dS_{t}= \mu S_{t} dt+S_{t} dW_{t},\qquad S_{0}=s>0,
	\end{equation*}
for all $t\in\left[0,1\right]$. Thus, setting $\rho\triangleq \exp\!\left\{-\mu W_{1}-\frac{\mu^{2}}{2}\right\}$, it is a well-known fact in the literature that the probability measure $\mathbb{Q}$ given by the Radon-Nikodym derivative $d\mathbb{Q}/d\mathbb{P}=\rho$ is the unique equivalent martingale measure, and that the process $\tilde{W}=\left\{\tilde{W}_{t}\textrm{; } 0\leq t \leq 1\right\}$ defined by
	\begin{equation*}
		\tilde{W}_{t}\triangleq W_{t}+\mu t
	\end{equation*}
is a $\mathbb{Q}$ Wiener process. Now take $A\triangleq \left\{\tilde{W}_{1}>0\right\}$. Clearly $\mathbb{Q}\!\left(A\right)=1/2$ and
	\begin{equation*}
		\mathbb{P}\!\left(A\right)=\mathbb{P}\!\left\{W_{1}+\mu>0\right\}=1-\mathbb{P}\!\left\{W_{1}\leq-\mu\right\}>1-\epsilon,
	\end{equation*}
	which then guarantees that $\mathbb{P}\!\left(A\right)^{\gamma}-\left[1-\mathbb{P}\!\left(A\right)\right]^{\delta}>0$, as intended, and the problem is ill-posed.
\end{Obs*}


\begin{remark}
	In \citet{he2011}, the authors introduce the concept of large-loss aversion degree (LLAD), defined as $\lim_{x\rightarrow +\infty} \frac{u_{-}\!\left(x\right)}{u_{+}\!\left(x\right)}$ (assuming that the limit exists), which can be interpreted as measuring ``the ratio between the pain of a substantial loss and the pleasure of a gain of the same magnitude'' \citep[Subsection~3.2]{he2011}. In our particular model, it is straightforward to see that
	\begin{equation*}
		\lim_{x\rightarrow +\infty} \frac{u_{-}\!\left(x\right)}{u_{+}\!\left(x\right)}=\lim_{x\rightarrow +\infty} \frac{x^{\beta}}{x^{\alpha}}=\left\{
			\begin{array}{ll}
				0, & \textrm{if } \alpha>\beta,\\
				1, & \textrm{if } \beta=\alpha,\\
				+\infty, & \textrm{if } \beta>\alpha,
			\end{array}
		\right.
	\end{equation*}
	so our Proposition~\ref{a>b} is in agreement with Theorem~1 in the above paper for one-step models. However, we will see that, unlike in the single-period model \citep[Corollary~4]{he2011}, in our continuous-time complete model the condition $\alpha<\beta$ alone is insufficient to ensure well-posedness.
\end{remark}


Let us now introduce the following auxiliary lemma, which will be used later.
\begin{lemma}\label{bd<1aux1}
	Let $X$ be a random variable such that $X \geq 0$ a.s. and $\mathbb{E}\!\left[X\right]=+\infty$. Then for each nonnegative real number $b$, there exists some $a=a\!\left(b\right) \in \left[\left.b,+\infty\right)\right.$ such that $b=\mathbb{E}\!\left[X\wedge a\right]$.
\end{lemma}
\begin{proof}
	The result follows from straightforward applications of Lebesgue's Monotone\index{Monotone Convergence Theorem} and Dominated Convergence Theorems\index{Dominated Convergence Theorem} (respectively MCT and DCT for short), as well as of the Intermediate Value Theorem, and the proof is relegated to the appendix.\qed
\end{proof}


By virtue of Proposition~\ref{a>b} above, it is now evident that we must impose $\alpha< \beta$ as a necessary condition for well-posedness. However, this is not enough to completely exclude ill-posedness, as shown by the next two propositions.
\begin{proposition}\label{bd<1}
	If $\frac{\beta}{\delta}<1$, then the problem~(\ref{eq:optport}) is ill-posed.
\end{proposition}
\begin{proof}
	Let us choose initial capital $x_{0}=0$. Given the hypothesis that $\frac{\beta}{\delta}<1$, there exists some $\chi$ such that $\frac{\beta}{\delta}<\chi<1$, and consequently we can choose $p \in \left(1,\frac{\delta \chi}{\beta}\right)$. Also, fix $0<\xi<\frac{\alpha}{\gamma}$. In particular, this implies that $\frac{\alpha}{\gamma \xi}>1$, and so there is $q>1$ such that $q \leq \frac{\alpha}{\gamma \xi}$.
	
	We define the nonnegative random variables
	\begin{equation*}
		Y\colonequals \left\{
			\begin{array}{ll}
				\frac{1}{U^{1/\xi}}, & \textrm{if } U<\frac{1}{2},\\
				0, & \textrm{if } U \geq \frac{1}{2},
			\end{array}
		\right.\quad
		\textrm{ and }\quad
		Z\colonequals \left\{
			\begin{array}{ll}
				0, & \textrm{if } U<\frac{1}{2},\\
				\frac{1}{\left(1-U\right)^{1/\chi}}, & \textrm{if } U \geq \frac{1}{2},
			\end{array}
		\right.
	\end{equation*}
	with $U \stackrel{_{\mathbb{Q}}}{\sim} \mathscr{U}\!\left(0,1\right)$ given in the proof of the previous proposition. Then, %
	since $\frac{1}{\chi}>1$, we obtain
	\begin{equation*}
		\mathbb{E}_{\mathbb{Q}}\!\left[Z\right]=\int_{0}^{1} \frac{1}{\left(1-u\right)^{1/\chi}} \bbOne_{\left[\left.\frac{1}{2},+\infty\right)\right.}\!\left(u\right) du=\int_{0}^{\frac{1}{2}} \frac{1}{u^{1/\chi}} du=+\infty.
	\end{equation*}
	In addition, using the reverse H\"older inequality\index{H\"older's inequality!reverse}, the monotonicity of the integral, and $\frac{\gamma q  \xi}{\alpha} \leq 1$, we conclude
	\begin{eqnarray*}
		\int_{0}^{+\infty} \mathbb{P}\!\left\{Y^{\alpha}>y\right\}^{\gamma} dy &=& \int_{0}^{+\infty} \mathbb{E}_{\mathbb{Q}}\!\left[\frac{1}{\rho} \bbOne_{\left\{Y^{\alpha}>y\right\}}\right]^{\gamma} dy\\
		& \geq & C_{1} \int_{0}^{+\infty} \mathbb{Q}\!\left\{Y^{\alpha}>y\right\}^{\gamma q} dy \geq C_{1} \int_{2^{\frac{\alpha}{\xi}}}^{+\infty} \frac{1}{y^{\frac{\gamma q  \xi}{\alpha}}} dy=+\infty,
	\end{eqnarray*}
	where $C_{1}\colonequals \mathbb{E}_{\mathbb{Q}}\!\left[\rho^{1/\left(q-1\right)}\right]^{\gamma \left(1-q\right)} \in \left(0,+\infty\right)$, and the last inequality follows from the fact that $\mathbb{Q}\!\left\{Y^{\alpha}>y\right\}=\mathbb{Q}\!\left\{U<\frac{1}{y^{\xi/\alpha}}\right\}$ for all $y \geq 2^{\alpha/\xi}$. Analogously, but this time by the conventional H\"older's inequality\index{H\"older's inequality}, we obtain that
	\begin{eqnarray*}
		\int_{0}^{+\infty} \mathbb{P}\!\left\{Z^{\beta}>y\right\}^{\delta} dy &\leq& C_{2} \int_{0}^{+\infty} \mathbb{Q}\!\left\{Z^{\beta}>y\right\}^{\frac{\delta}{p}} dy\\
		&\leq& 2^{\frac{\beta}{\chi}} C_{2}+C_{2} \int_{2^{\frac{\beta}{\delta}}}^{+\infty} \frac{1}{y^{\frac{\delta \chi}{\beta p}}} dy<+\infty,
	\end{eqnarray*}
	with $C_{2}\colonequals \mathbb{E}_{\mathbb{Q}}\!\left[\frac{1}{\rho^{p/\left(p-1\right)}}\right]^{\delta \frac{p-1}{p}} \in \left(0,+\infty\right)$, since $\frac{\delta \chi}{\beta p}>1$ and $\mathbb{Q}\!\left\{Z^{\beta}>y\right\}=\mathbb{Q}\!\left\{U>1-\frac{1}{y^{\chi/\beta}}\right\}$ for all $y \geq 2^{\beta/\chi}$.
	
	Now we set $Y_{n}\colonequals Y\wedge n$ for each $n \in \mathbb{N}$. Clearly each $Y_{n}$ is nonnegative, and given that $\mathbb{Q}\!\left\{Y>0\right\}=\mathbb{Q}\!\left\{U<\frac{1}{2}\right\}=1/2$, we conclude that $\mathbb{E}_{\mathbb{Q}}\!\left[Y_{n}\right]>0$
. We also note that $\mathbb{E}_{\mathbb{Q}}\!\left[Y_{n}\right]\leq n <+\infty$, hence it follows from Lemma~\ref{bd<1aux1} that $\mathbb{E}_{\mathbb{Q}}\!\left[Y_{n}\right]=\mathbb{E}_{\mathbb{Q}}\!\left[Z\wedge a_{n}\right]$ for some strictly positive real number $a_{n}$. So let us define $Z_{n}\colonequals Z\wedge a_{n}$ for every $n$. Then each $Z_{n}$ is a nonnegative random variable satisfying $V_{-}\!\left(Z_{n}\right)\leq V_{-}\!\left(Z\right)<+\infty$.
	
	Finally, we consider the sequence of $\sigma\!\left(\rho\right)$-measurable random variables $\left\{X_{n}\right\}_{n \in \mathbb{N}}$, with $X_{n}\colonequals Y_{n}-Z_{n}$. It is clear, by the way it was constructed, that $\mathbb{E}_{\mathbb{Q}}\!\left[X_{n}\right]=0$, $X_{n}^{+}=Y_{n}$, $X_{n}^{-}=Z_{n}$, and $X_{n}\geq -a_{n}$, for all $n$. Also, $V_{-}\!\left(X_{n}^{-}\right)=V_{-}\!\left(Z_{n}\right)<+\infty$, so $X_{n}$ is feasible for~\eqref{eq:optport}. Consequently, the problem is ill-posed because we get
	\begin{equation*}
		V\!\left(X_{n}\right)\geq V_{+}\!\left(Y_{n}\right)-V_{-}\!\left(Z\right) \xrightarrow[n\rightarrow +\infty]{} V_{+}\!\left(Y\right)-V_{-}\!\left(Z\right)=+\infty
	\end{equation*}
using monotone convergence, since $V_{-}\!\left(Z\right)$ is finite.\qed
\end{proof}


\begin{proposition}\label{ag>1}
	If $\frac{\alpha}{\gamma}>1$, then the problem is ill-posed.
\end{proposition}
\begin{proof}
	As before, let the initial capital $x_{0}$ equal zero and $U$ be the uniform random variable defined in the proof of Proposition~\ref{a>b}. Denoting by $Y$ the nonnegative random variable given by
	\begin{equation*}
		Y\colonequals \left\{
			\begin{array}{ll}
				\frac{1}{U^{1/\xi}}, & \textrm{if } U<\frac{1}{2},\\
				0, & \textrm{if } U \geq \frac{1}{2},
			\end{array}
		\right.
	\end{equation*}
where $\xi$ is chosen in such a way that $1<\xi<\frac{\alpha}{\gamma}$, we see that $\mathbb{E}_{\mathbb{Q}}\!\left[Y\right]=\int_{0}^{\frac{1}{2}} \frac{1}{u^{1/\xi}} du<+\infty$. Moreover, applying the reverse H\"older inequality as in the proof of Proposition~\ref{bd<1} with $p=\frac{\alpha}{\gamma \xi}>1$, yields
	\begin{equation*}
		V_{+}\!\left(Y\right)=\int_{0}^{+\infty} \mathbb{P}\!\left\{Y^{\alpha}>y\right\}^{\gamma} dy \geq C \int_{2^{\frac{\alpha}{\xi}}}^{+\infty} \frac{1}{y} dy=+\infty,
	\end{equation*}
	where the positive constant $C$ (depending on $p$) equals $\mathbb{E}_{\mathbb{Q}}\!\left[\rho^{1/\left(p-1\right)}\right]^{\gamma\left(1-p\right)}$. Finally, for each $n \in \mathbb{N}$, we define the nonnegative random variable $Y_{n}\colonequals Y\wedge n$ and we set $C_{n}\colonequals \mathbb{E}_{\mathbb{Q}}\!\left[Y_{n}\right] \in \left(\left.0,n\right]\right.$. Then, for the nonnegative random variable $Z_{n}=2\,C_{n}\,\bbOne_{\left\{U \geq \frac{1}{2}\right\}}$, we have that $V_{-}\!\left(Z_{n}\right)=\left(2\,C_{n}\right)^{\beta} \mathbb{P}\!\left\{U \geq \frac{1}{2}\right\}^{\delta}$, for all $n$. We note further that $\mathbb{P}\!\left\{U \geq \frac{1}{2}\right\}>0$ (we recall that $\mathbb{P}$ is equivalent to $\mathbb{Q}$) and $C_{n} \xrightarrow[n\rightarrow +\infty]{} \mathbb{E}_{\mathbb{Q}}\!\left[Y\right]$ (by the Monotone Convergence Theorem).
	
	Now take $X_{n}\colonequals Y_{n}-Z_{n}$. It is clear that $X_{n}^{+}=Y_{n}$ and $X_{n}^{-}=Z_{n}$. Additionally, being $\sigma\!\left(\rho\right)$-measurable, bounded from below by $-2\,C_{n}$, and satisfying $\mathbb{E}_{\mathbb{Q}}\!\left[X_{n}\right]=\mathbb{E}_{\mathbb{Q}}\!\left[Y_{n}\right]-2\,C_{n}\,\mathbb{Q}\!\left\{U \geq \frac{1}{2}\right\}=0$, as well as $V_{-}\!\left(X_{n}^{-}\right)<+\infty$, each $X_{n}$ is feasible for the problem. However, because $\mathbb{E}_{\mathbb{Q}}\!\left[Y\right]$ is finite,
	\begin{eqnarray*}
		V\!\left(X_{n}\right)&=& V_{+}\!\left(Y_{n}\right)-\left(2\,C_{n}\right)^{\beta} \mathbb{P}\!\left\{U \geq \frac{1}{2}\right\}^{\delta}\\
		& \rightarrow& V_{+}\!\left(Y\right)-\left(2\,\mathbb{E}_{\mathbb{Q}}\!\left[Y\right]\right)^{\beta} \mathbb{P}\!\left\{U \geq\frac{1}{2}\right\}^{\delta}=+\infty
	\end{eqnarray*}
as $n \rightarrow +\infty$, which completes the proof.\qed
\end{proof}


The preceding results can then be summarised in the form of a theorem.
\begin{theorem}\label{necesscond}
	The problem~(\ref{eq:optport}) is well-posed only if
	\begin{equation}\label{eq:necesscond}
		\alpha<\beta \text{\quad and \quad} \frac{\alpha}{\gamma}\leq 1\leq \frac{\beta}{\delta}.
	\end{equation}
\end{theorem}

\begin{remark}
	An inspection of the proofs above reveals that, if \eqref{eq:necesscond} is not satisfied, then the maximisation problem is ill-posed even over the set of feasible strategies with bounded terminal values.
\end{remark}

In conclusion, for the well-posedness of~\eqref{eq:optport}, it is necessary to have both $\alpha< \beta$ and $\frac{\alpha}{\gamma} \leq 1 \leq \frac{\beta}{\delta}$, which are conditions that are very easy to be checked. We note further that they admit an intuitive interpretation. As a matter of fact, the first condition means that the \emph{S}-shaped utility function is steeper in the negative domain than in the positive one, or economically speaking, that losses loom larger than corresponding gains, indicating loss aversion. As for the inequalities $\alpha/\gamma \leq 1 \leq \beta/\delta$, they reflect the fact that the investor's risk preferences and perceptions, both on losses and on gains, have to be well-adjusted, in the sense that, for instance, the distortion on losses cannot override the pain of a loss.

\begin{remark}
	In the particular case where $\delta=1$ (no distortion on the negative side), it follows from Theorem~\ref{necesscond} that the problem is ill-posed for all $\beta \in \left(0,1\right)$. Hence, a probability distortion on losses is a necessary condition for the well-posedness of~(\ref{eq:optport}), which is in line with Theorem~3.2 in the paper for \citet{jin2008} in the context of a continuous complete market. We stress, however, that under the assumptions of Theorem~4.4 in \citet{carassus2011}, and regardless of the fact that there is a probability distortion on losses or not, the optimal portfolio problem in the multiperiod incomplete financial market model under consideration can be well-posed. Also, \citet[Remark~3.1]{jin2008} notice that the problem in \citet{berkelaar04} is well-posed though no probability distortion is considered. This is because the wealth process is required to be nonnegative in \citet{berkelaar04}.
	
	Summing up these arguments, we are led to believe that ill-posedness in our continuous-time model is likely to be due precisely to the richness of attainable payoffs \citep[Remark~4.10]{carassus2011}, as well as to the absence of any constraints on the wealth process.
\end{remark}

The rest of this section is devoted to the proof of the (almost) reciprocal of Theorem~\ref{necesscond}, and thus to showing that~(\ref{eq:necesscond}) above is essentially sharp. This requires that we introduce some auxiliary lemmata first, which are proved in the appendix.
%
%
\begin{lemma}\label{EleqL}
	If $a$, $b$ and $s$ are strictly positive real numbers satisfying $\frac{b}{s a}>1$, then there exists a strictly positive real constant $D$ (depending on $a$, $b$ and $s$) such that
	\begin{equation}\label{eq:EleqL}
		\mathbb{E}_{\mathbb{P}}\!\left[X^{s}\right]\leq 1+D\left(\int_{0}^{\infty} \mathbb{P}\!\left\{X^{b}>y\right\}^{a} dy\right)^{\frac{1}{a}},
	\end{equation}
	for all nonnegative real-valued random variables $X$.
\end{lemma}


\begin{lemma}\label{Lemeta}
	Let $\alpha<\beta$ and $\frac{\alpha}{\gamma}<1<\frac{\beta}{\delta}$. Then there is some $\eta>0$ satisfying $\eta<\beta$, $\alpha<\eta$ and $\delta<\eta$, and there exist strictly positive constants $L_{1}$ and $L_{2}$ such that
	\begin{equation}\label{eq:Lemeta}
		\int_{0}^{+\infty} \mathbb{P}\!\left\{\left(X^{+}\right)^{\alpha}>y\right\}^{\gamma} dy \leq L_{1}+L_{2}\int_{0}^{+\infty} \mathbb{P}\!\left\{\left(X^{-}\right)^{\eta}>y\right\}^{\delta} dy,
	\end{equation}
	for all random variables $X$ with $\mathbb{E}_{\mathbb{Q}}\!\left[X\right]=x_{0}$.
\end{lemma}


\begin{lemma}\label{L1L2}
	Let $a$, $b$ and $s$ be strictly positive real numbers such that $s<a<b$ and $s\leq 1$. Then there exist $0<\zeta<1$ and strictly positive constants $R_{1}$, $R_{2}$ such that
	\begin{equation}
		\int_{0}^{+\infty} \mathbb{P}\!\left\{X^{a}>y\right\}^{s} dy \leq R_{1}+R_{2}\left(\int_{0}^{+\infty} \mathbb{P}\!\left\{X^{b}>y\right\}^{s} dy\right)^{\zeta}
	\end{equation}
	for all nonnegative random variables $X$. In particular, this implies that we have $\int_{0}^{+\infty} \mathbb{P}\!\left\{X_{n}^{b}>y\right\}^{s} dy \xrightarrow[n\rightarrow +\infty]{} +\infty$ whenever $\int_{0}^{+\infty} \mathbb{P}\!\left\{X_{n}^{a}>y\right\}^{s} dy \xrightarrow[n\rightarrow +\infty]{} +\infty$, for any sequence $\left\{X_{n}\right\}_{n \in \mathbb{N}}$ of nonnegative random variables.
\end{lemma}


Hence, we manage to provide the following sufficient condition for well-posedness.
\begin{theorem}\label{suffcond}
	Suppose that %
	Assumption~\ref{as:CDFW} holds and
	\begin{equation}\label{eq:suffcond}
		\alpha<\beta \textrm{\quad and \quad} \frac{\alpha}{\gamma}<1<\frac{\beta}{\gamma}.
	\end{equation}
	Then the optimisation problem~\eqref{eq:optport} is well-posed.
\end{theorem}
\begin{proof}
	The proof is by contradiction. Let us suppose that the optimisation problem is ill-posed, that is, $\sup_{\phi \in \mathscr{A}\!\left(x_{0}\right)} V\!\left(\Pi^{\phi}_{T}\right)=+\infty$. Then we can take a sequence of trading strategies $\left\{\phi^{(n)}\text{; } n \in \mathbb{N}\right\}$ in the feasible set $\mathscr{A}\!\left(x_{0}\right)$ such that $\lim_{n\rightarrow +\infty} V\!\left(X_{n}\right)=+\infty$, where for each $n$ the random variable $X_{n}$ denotes the value of the $n$-th portfolio at the terminal time $T$, $X_{n}=\Pi^{\phi^{(n)}}_{T}$.
	
	Since we have $V_{+}\!\left(X_{n}^{+}\right)\geq V\!\left(X_{n}\right)$ for every $n \in \mathbb{N}$, we can deduce that $V_{+}\!\left(X_{n}^{+}\right)\rightarrow +\infty$ as $n \rightarrow +\infty$. Thus, it follows from Lemma~\ref{Lemeta} that
	\begin{equation*}
		\lim_{n\rightarrow +\infty} \int_{0}^{+\infty} \mathbb{P}\!\left\{\left(X_{n}^{-}\right)^{\eta}>y\right\}^{\delta} dy =+\infty
	\end{equation*}
	for some $\eta$ satisfying $\eta<\beta$, $\alpha<\eta$ and $\delta<\eta$. Consequently, we can apply Lemma~\ref{L1L2} to conclude that also $\lim_{n\rightarrow +\infty} V_{-}\!\left(X_{n}^{-}\right)=+\infty$.
	
	Therefore, again using Lemma~\ref{Lemeta} and Lemma~\ref{L1L2} (and recalling that $0<\zeta<1$),
	\begin{eqnarray*}
		V\!\left(X_{n}\right)&\leq& L_{1}+L_{2} \int_{0}^{+\infty} \mathbb{P}\!\left\{\left(X_{n}^{-}\right)^{\eta}>y\right\}^{\delta} dy-V_{-}\!\left(X_{n}^{-}\right)\\
		&\leq& \left(R_{1}+L_{2} R_{1}\right)+L_{2} R_{2}\left[V_{-}\!\left(X_{n}^{-}\right)\right]^{\zeta}-V_{-}\!\left(X_{n}^{-}\right) \xrightarrow[n\rightarrow +\infty]{} -\infty,
	\end{eqnarray*}
	which is absurd. Hence, as claimed, the problem is well-posed.\qed
\end{proof}

\begin{remark}
It is worth comparing Theorem~\ref{suffcond} to a result of
\citet{carassus2011}. In a discrete-time,
multiperiod market model,
well-posedness is obtained in \citet{carassus2011} whenever 
\begin{equation}\label{prima}
\alpha/\gamma<\beta,
\end{equation} 
modulo some integrability conditions related to the price process.
One can check, using Proposition 7.1 of \citet{rs05},
that the integrability conditions of \citet{carassus2011} imply the existence
of a risk-neutral measure $\mathbb{Q}\sim\mathbb{P}$ with $d\mathbb{Q}/d\mathbb{P}\text{, }
d\mathbb{P}/d\mathbb{Q}\in \mathscr{W}$, hence Theorem~\ref{suffcond} implies
well-posedness also in the case 
\begin{equation}\label{secunda}
\alpha<\beta, \quad \alpha/\gamma<1<\beta/\delta.
\end{equation}
Note that \eqref{prima} and \eqref{secunda} are incomparable 
conditions (none of them implies the other one). It is also worth
emphasizing that the domains of optimization are different in \citet{carassus2011}
and in the present paper. Hence, in the discrete-time multiperiod case,
our Theorem~\ref{suffcond} complements (but does not subsume) the corresponding
results of \citet{carassus2011}.
\end{remark}


\section{Attainability}\label{ch:Existence}

We shall assume, as we did in Section~\ref{piecewisepower}, that the utility functions and the probability distortions are of the form given by~\eqref{eq:poweru} and \eqref{eq:powerw}. Moreover, conditions $\alpha<\beta$ and $\alpha/\gamma<1<\beta/\delta$ will also be in force throughout this section. Theorem~\ref{suffcond} then ensures that the problem~\eqref{eq:optport} is well-posed. However, even if the problem is well-posed, it may still happen that an optimal solution does not exist. Considering this, we state the following.
\begin{definition}
	A well-posed optimisation problem is said to be \emph{attainable}\index{attainability} if it admits an optimal solution.
\end{definition}

Our aim is therefore to investigate the existence of an optimal trading strategy, that is, to determine whether the supremum is indeed a maximum or not. As in the previous section, our ideas are essentially inspired in the work of \citet{carassus2011}, in particular we follow the proof of Theorem~5.6 closely.

We begin by noticing that, since the supremum of the maximisation problem~\eqref{eq:optport} is finite, say
\begin{equation}\label{eq:Vstar}
	\sup_{\phi \in \mathscr{A}\!\left(x_{0}\right)} V\!\left(\Pi^{\phi}_{T}\right)=V^{*},
\end{equation}
for some $V^{*}<+\infty$, we can take a sequence of feasible portfolios $\left\{\phi^{(n)}\text{; } n \in \mathbb{N}\right\}$ such that $\lim_{n\rightarrow +\infty} V\!\left(\Pi^{\phi^{(n)}}_{T}\right)=V^{*}$. For the sake of convenience of writing, and as in the proof of Theorem~\ref{suffcond}, we shall henceforth denote by $X_{n}$ the terminal wealth of the $n$-th portfolio $\phi^{(n)}$. Moreover, for each $n \in \mathbb{N}$, we recall that the law\index{law} of $X_{n}$ with respect to $\mathbb{P}$ is defined by $\mathbb{P}_{X_{n}}\!\left(A\right)\colonequals \mathbb{P}\!\left\{X_{n} \in A\right\}$\nomenclature[P]{$\mathbb{P}_{X}$}{Law of $X$ with respect to $\mathbb{P}$} for all Borel sets $A \subseteq \mathbb{R}$.


We then introduce the following relevant technical lemma, which is proved in the appendix.
\begin{lemma}\label{supEtau}
	There exists $\tau \in \left(0,1\right)$ such that $\sup_{n \in \mathbb{N}} \mathbb{E}_{\mathbb{P}}\!\left[\left|X_{n}\right|^{\tau}\right]<+\infty$.
\end{lemma}

From this, it is immediate to derive the following result (whose proof can also be found in the appendix).
\begin{corollary}\label{tight1}
	The sequence of the laws of the random variables $X_{n}$ is tight\index{tightness}, that is, for every $\epsilon>0$ there is a compact subset $K=K\!\left(\epsilon\right)$ of $\mathbb{R}$ such that $\mathbb{P}\!\left(X_{n} \in K^{c}\right)<\epsilon$ for all $n \in \mathbb{N}$.
\end{corollary}

Now let $\nu_{n}$\nomenclature[n]{$\nu_{n}$}{Joint law of the random vector $\left(\rho,X_{n}\right)$} denote the joint law of the random vector $\left(\rho,X_{n}\right)$. As a consequence of the preceding corollary, along with Ulam's theorem\index{Ulam's theorem} that every finite measure on a Polish space\index{Polish space} is tight%
, we trivially obtain the next important lemma.
\begin{lemma}\label{tight2}
	The sequence $\left\{\nu_{n}\text{; }n \in \mathbb{N}\right\}$ is also tight.\qed
\end{lemma}

Then, by \citet{prokhorov56} theorem, the family $\left\{\nu_{n}\text{; }n \in \mathbb{N}\right\}$ is weakly compact, that is, we can extract from $\left\{\nu_{n}\text{; }n \in \mathbb{N}\right\}$ a weakly convergent subsequence $\left\{\nu_{n_{k}}\text{; } k \in \mathbb{N}\right\}$, and we write $\nu_{n_{k}} \stackrel{w}{\longrightarrow} \nu$ for some probability measure $\nu$.

Now, since $\nu$ is a probability measure on the product space $\left(\mathbb{R}^{2},\mathscr{B}\!\left(\mathbb{R}^{2}\right)\right)$ and $\left(\mathbb{R},\mathscr{B}\!\left(\mathbb{R}\right)\right)$ is a standard space\index{standard space}, by the disintegration theorem\index{disintegration theorem} there exists a probability measure $\nu_{*}$ on $\mathbb{R}$ and a transition probability kernel $K$\index{transition probability kernel} on $\left(\mathbb{R},\mathscr{B}\!\left(\mathbb{R}\right)\right)$ such that $\nu\!\left(A_{1}\times A_{2}\right)=\int_{A_{1}} K\!\left(x,A_{2}\right)d\nu_{*}\left(x\right)$ for all $A_{1},A_{2} \in \mathscr{B}\!\left(\mathbb{R}\right)$%
.%
\footnote{Here $\mathscr{B}\!\left(X\right)$ denotes the Borel $\sigma$-algebra on the topological space $X$. A mapping $K$ from $\mathbb{R}\times \mathscr{B}\!\left(\mathbb{R}\right)$ into $\left[0,+\infty\right]$ is called a transition probability kernel on $\left(\mathbb{R},\mathscr{B}\!\left(\mathbb{R}\right)\right)$ if the mapping $x \mapsto K\!\left(x,B\right)$ is measurable for every set $B \in \mathscr{B}\!\left(\mathbb{R}\right)$, and the mapping $B \mapsto K\!\left(x,B\right)$ is a probability measure for every $x \in \mathbb{R}$.} %
We call $\nu_{*}$ the marginal of $\nu$ with respect to its first coordinate. Analogously, for each $n$, let $\lambda_{n}$ denote the probability measure on $\mathbb{R}$ that represents the marginal of $\nu_{n}$ with respect to its first coordinate. Clearly, for every $n$, $\lambda_{n}\!\left(A\right)=\mathbb{P}\!\left(\rho \in A\right)$ for all Borel sets $A\subseteq R$. Furthermore we have the following obvious result.
\begin{lemma}\label{lawrho}
	For all $A \in \mathscr{B}\!\left(\mathbb{R}\right)$, $\nu_{*}\!\left(A\right)=\mathbb{P}_{\rho}\!\left(A\right)$, where $\mathbb{P}_{\rho}$ is the law of $\rho$ under $\mathbb{P}$.\qed
\end{lemma}


Also, we present a crucial auxiliary lemma, which is essentially Lemma~8.5 in \citet{carassus2011}.
\begin{lemma}\label{Glawnu}
	Suppose that Assumption~\ref{as:Ustar} holds. Then there exists a measurable function $G:\mathbb{R}\times \left(0,1\right)\rightarrow \mathbb{R}$ so that the random vector $\left(\rho,G\!\left(\rho,U_{*}\right)\right)$ has law $\nu$.\qed
\end{lemma}


So we are finally in the position to state the main result of this section.
\begin{theorem}\label{Theo:Exist}
	There exists an optimal trading strategy.
\end{theorem}

\begin{proof}
	Let us set $X_{*} \colonequals G\!\left(\rho,U_{*}\right)$, where $G$ is the measurable function given by Lemma~\ref{Glawnu}. Clearly the random variable $X_{*}$ is $\sigma\!\left(\rho,U_{*}\right)$-measurable.
	
	Moreover, from the discussion above, it is straightforward to check that the subsequence of random variables $\left\{X_{n_{k}}\right\}_{k \in \mathbb{N}}$ converges in distribution to $X_{*}$ as $k \rightarrow +\infty$, and we write $X_{n_{k}} \stackrel{\mathscr{D}}{\longrightarrow} X_{*}$. Then, since the maximum and $u_{+}$ are continuous functions, by the mapping theorem we have that $\left\{u_{+}\!\left(X_{n_{k}}^{+}\right)\right\}_{k \in \mathbb{N}}$ also converges in distribution to $u_{+}\!\left(X_{*}^{+}\right)$. Hence, $\lim_{k \in \mathbb{N}} \mathbb{P}\!\left\{u_{+}\!\left(X_{n_{k}}^{+}\right)>y\right\}=\mathbb{P}\!\left\{u_{+}\!\left(X_{*}^{+}\right)>y\right\}$ for every $y \in \mathbb{R}$ at which the CDF of $u_{+}\!\left(X_{*}^{+}\right)$ is continuous (and we recall that any non-decreasing function has at most countably many discontinuities). Analogously we conclude that $\mathbb{P}\!\left\{u_{-}\!\left(X_{n_{k}}^{-}\right)>y\right\} \rightarrow \mathbb{P}\!\left\{u_{-}\!\left(X_{*}^{-}\right)
>y\right\}$ as $k \rightarrow +\infty$ for all $y$ outside a countable set.
	
	At this point, we divide the proof into three steps.
	\begin{enumerate}[label=\emph{(\roman*)}]
		\item
		We start by showing that $V_{\pm}\!\left(X_{*}^{\pm}\right)<+\infty$. In fact, given that the distortion functions are continuous, it is obvious that $w_{\pm}\!\left(\mathbb{P}\!\left\{u_{\pm}\!\left(X_{n_{k}}^{\pm}\right)>y\right\}\right) \xrightarrow[k\rightarrow +\infty]{} w_{\pm}\!\left(\mathbb{P}\!\left\{u_{\pm}\!\left(X_{*}^{\pm}\right)>y\right\}\right)$ for Lebesgue a.e.\ $y$. Thus, applying Fatou's lemma\index{Fatou's lemma} (we recall that the distortion functions are nonnegative) we get
		\begin{eqnarray*}
			V_{\pm}\!\left(X_{*}^{\pm}\right) &=& \int_{0}^{+\infty} w_{\pm}\!\left(\mathbb{P}\!\left\{u_{\pm}\!\left(X_{*}^{\pm}\right)>y\right\}\right) dy\\
			&\leq& \liminf_{k \in \mathbb{N}} \int_{0}^{+\infty} w_{\pm}\!\left(\mathbb{P}\!\left\{u_{\pm}\!\left(X_{n_{k}}^{\pm}\right)>y\right\}\right) dy= \liminf_{k \in \mathbb{N}} V_{\pm}\!\left(X_{n_{k}}^{\pm}\right).
		\end{eqnarray*}
		But $\liminf_{k \in \mathbb{N}} V_{\pm}\!\left(X_{n_{k}}^{\pm}\right)\leq \sup_{k \in \mathbb{N}} V_{\pm}\!\left(X_{n_{k}}^{\pm}\right)\leq \sup_{n \in \mathbb{N}} V_{\pm}\!\left(X_{n}^{\pm}\right)$, and we know from the proof of Lemma~\ref{supEtau} that $\sup_{n \in \mathbb{N}} V_{\pm}\!\left(X_{n}^{\pm}\right)<+\infty$, so we have the intended result.
		
		\item
		Secondly, we prove that the inequality $V\!\left(X_{*}\right)\geq V^{*}$ holds. We already know, from the previous step, that $V_{-}\!\left(X_{*}^{-}\right)\leq \liminf_{k \in \mathbb{N}} V_{-}\!\left(X_{n_{k}}^{-}\right)$. We note further that, by the proof of Lemma~\ref{supEtau}, $\sup_{k \in \mathbb{N}} \mathbb{E}_{\mathbb{P}}\!\left[\left(X_{n_{k}}^{+}\right)^{\alpha \lambda}\right]\leq \sup_{n \in \mathbb{N}} \mathbb{E}_{\mathbb{P}}\!\left[\left(X_{n}^{+}\right)^{\alpha \lambda}\right]<+\infty$, for some $\lambda>0$ such that $\alpha \lambda < 1 < \gamma \lambda$. Therefore, defining $g\!\left(y\right)=1$ for $y\in \left[0,1\right]$ and $g\!\left(y\right)=\frac{\left(\sup_{n \in \mathbb{N}} \mathbb{E}_{\mathbb{P}}\!\left[\left(X_{n}^{+}\right)^{\alpha \lambda}\right]\right)^{\gamma}}{y^{\gamma \lambda}}$ for $y>1$, we conclude that $g$ is an integrable function on $\left[\left.0,+\infty\right)\right.$. Moreover, it follows from Chebyshev's inequality\index{Chebyshev's inequality} that $w_{+}\!\left(\
mathbb{P}\!\left\{u_{+}\!\left(X_{n_{k}}^{+}\right)>y\right\}\right)\leq g\!\left(y\right)$ for all $y\geq 0$ and for all $k \in \mathbb{N}$. Hence, we can apply the reverse Fatou lemma\index{Fatou's lemma!reverse} to obtain
		\begin{equation*}
			V_{+}\!\left(X_{*}^{+}\right)\geq \limsup_{k \in \mathbb{N}} \int_{0}^{+\infty} w_{+}\!\left(\mathbb{P}\!\left\{u_{+}\!\left(X_{n_{k}}^{+}\right)>y\right\}\right) dy=\limsup_{k \in \mathbb{N}} V_{+}\!\left(X_{n_{k}}^{+}\right).
		\end{equation*}
		Combining the previous inequalities then yields
		\begin{eqnarray*}
			V\!\left(X_{*}\right) &=& V_{+}\!\left(X_{*}^{+}\right)-V_{-}\!\left(X_{*}^{-}\right)\geq \limsup_{k \in \mathbb{N}} V_{+}\!\left(X_{n_{k}}^{+}\right)-\liminf_{k \in \mathbb{N}} V_{-}\!\left(X_{n_{k}}^{-}\right)\\
			&=& \limsup_{k \in \mathbb{N}} V_{+}\!\left(X_{n_{k}}^{+}\right)+\limsup_{k \in \mathbb{N}} \left\{-V_{-}\!\left(X_{n_{k}}^{-}\right)\right\}\\
			&\geq & \limsup_{k \in \mathbb{N}} \left\{V_{+}\!\left(X_{n_{k}}^{+}\right)-V_{-}\!\left(X_{n_{k}}^{-}\right)\right\}=V^{*},
		\end{eqnarray*}
		where the last inequality is due to the subadditivity of the $\limsup$.
		
		\item
		Lastly, we check that $\mathbb{E}_{\mathbb{Q}}\!\left[ X_{*} \right]\leq x_{0}$. To see this, we start by noting that, since 
$\left(\rho,X_{n_{k}}\right) \stackrel{\mathscr{D}}{\longrightarrow} \left(\rho,X_{*}\right)$, 
again by applying the mapping theorem we have $\rho X_{n_{k}} \stackrel{\mathscr{D}}{\longrightarrow} \rho X_{*}$. Thus, we can use Skorohod's theorem\index{Skorohod's theorem} \citep[see e.g.][Theorem~2.2.2, p.~23]{borkar95} to find real-valued random variables $Y$ and $Y_{k}$, with $k \in \mathbb{N}$, on some auxiliary probability space $\left(\hat{\Omega},\hat{\mathscr{F}},\hat{\mathbb{Q}}\right)$, such that each $Y_{k}$ has the same law as $\rho X_{n_{k}}$, $Y$ has the same law as $\rho X_{*}$, and $Y_{k} \xrightarrow[k\rightarrow +\infty]{} Y$ $\hat{\mathbb{Q}}$ a.s.. It is then clear that $\mathbb{E}_{\mathbb{Q}}\!\left[X_{*}\right]=\mathbb{E}_{\mathbb{P}}\!\left[\rho X_{*}\right]=\mathbb{E}_{\hat{\mathbb{Q}}}\!\left[Y\right]$ holds, as well as $\mathbb{E}_{\mathbb{Q}}\!\left[X_{n_{k}}\right]=\mathbb{E}_{\mathbb{P}}\!\left[\rho X_{n_{k}}\right]=\mathbb{E}_{\hat{\mathbb{Q}}}\!\left[Y_{k}\right]$ for every $k \in \mathbb{N}$. Now, we know from the proof of Lemma~\ref{supEtau} that $\sup_{k \in \mathbb{N}} \mathbb{E}_{\mathbb{P}}\!\left[\left(X_{n_{k}}^{-}\right)^{\xi}\right]<+\infty$, for some $1<\xi<\frac{\beta}{\delta}$. Consequently, we can choose $\vartheta>1$ such that $\vartheta<\xi$, and so using H\"older's inequality we obtain
		\begin{equation*}
			\mathbb{E}_{\hat{\mathbb{Q}}}\!\left[\left(Y_{k}^{-}\right)^{\vartheta}\right]=\mathbb{E}_{\mathbb{P}}\!\left[\left(\rho X_{n_{k}}^{-}\right)^{\vartheta}\right]\leq \mathbb{E}_{\mathbb{P}}\!\left[\rho^{\frac{\vartheta \xi}{\xi-\vartheta}}\right]^{\frac{\xi-\vartheta}{\xi}} \mathbb{E}_{\mathbb{P}}\!\left[\left(X_{n_{k}}^{-}\right)^{\xi}\right]^{\frac{\vartheta}{\xi}},
		\end{equation*}
		for every $k \in \mathbb{N}$, which implies that $\sup_{k\in \mathbb{N}} \mathbb{E}_{\hat{\mathbb{Q}}}\!\left[\left(Y_{k}^{-}\right)^{\vartheta}\right]<+\infty$. Hence, by de la Vall\'ee-Poussin theorem\index{de la Vall\'ee-Poussin theorem}, the family $\left\{Y_{k}^{-}\right\}_{k \in \mathbb{N}}$ is uniformly integrable and thus, being also integrable, it holds that
		\begin{equation*}
			\lim_{k \in \mathbb{N}} \mathbb{E}_{\mathbb{\hat{Q}}}\!\left[Y_{k}^{-}\right]=\mathbb{E}_{\mathbb{\hat{Q}}}\!\left[Y^{-}\right]<+\infty.
		\end{equation*}
		Furthermore, using Fatou's lemma we easily get the inequality $\mathbb{E}_{\mathbb{\hat{Q}}}\!\left[Y^{+}\right]\leq \liminf_{k \in \mathbb{N}} \mathbb{E}_{\mathbb{\hat{Q}}}\!\left[Y_{k}^{+}\right]$. Combining all the results above finally yields
		\begin{eqnarray*}
			\mathbb{E}_{\mathbb{Q}}\!\left[X_{*}\right] &=& \mathbb{E}_{\hat{\mathbb{Q}}}\!\left[Y\right]\leq \liminf_{k \in \mathbb{N}} \mathbb{E}_{\mathbb{\hat{Q}}}\!\left[Y_{k}^{+}\right]-\lim_{k \in \mathbb{N}} \mathbb{E}_{\mathbb{\hat{Q}}}\!\left[Y_{k}^{-}\right]\\
			&\leq& \liminf_{k \in \mathbb{N}} \mathbb{E}_{\mathbb{\hat{Q}}}\!\left[Y_{k}\right]=\liminf_{k \in \mathbb{N}} \mathbb{E}_{\mathbb{Q}}\!\left[X_{n_{k}}\right]=x_{0},
		\end{eqnarray*}
		where the last inequality is a consequence of the fact that the $\liminf$ satisfies superadditivity.
	\end{enumerate}

	It is now straightforward to check that
	\begin{equation*}
		\mathbb{E}_{\mathbb{Q}}\!\left[\left|X_{*}\right|\right]=\mathbb{E}_{\hat{\mathbb{Q}}}\!\left[\left|Y\right|\right]=\mathbb{E}_{\hat{\mathbb{Q}}}\!\left[Y\right]+2\,\mathbb{E}_{\hat{\mathbb{Q}}}\!\left[Y^{-}\right]<+\infty,
	\end{equation*}
	that is, $X_{*} \in L^{1}\!\left(\mathbb{Q}\right)$. Hence, by Assumption~\ref{as:kindcompl}, $X_{*}$ admits a replicating portfolio $\phi^{*} \in \Psi$.
	
	We also know, by part~(i), that $V_{\pm}\!\left(\left[\Pi^{\phi^{*}}_{T}\right]^{\pm}\right)<+\infty$. Therefore, if the equality $\mathbb{E}_{\mathbb{Q}}\!\left[X_{*}\right]=x_{0}$ is satisfied, then $\phi^{*}$ is in $\mathscr{A}\!\left(x_{0}\right)$ and $V^{*}\leq V\!\left(\Pi^{\phi^{*}}_{T}\right)\leq \sup_{\phi \in \mathscr{A}\!\left(x_{0}\right)} V\!\left(\Pi^{\phi}_{T}\right)=V^{*}$, hence we are done. So let us suppose that $\mathbb{E}_{\mathbb{Q}}\!\left[X_{*}\right]<x_{0}$. We can thus define the $\mathscr{F}_{T}$-measurable random variable $Z_{*}\colonequals X_{*}+c$, where the constant $c$ is given by $c \colonequals x_{0}-\mathbb{E}_{\mathbb{Q}}\!\left[X_{*}\right]>0$. Then it is immediate to see that $Z_{*}$ is also replicable by a portfolio $\pi^{*} \in \Psi$ such that $\mathbb{E}_{\mathbb{Q}}\!\left[\Pi^{\pi^{*}}_{T}\right]=x_{0}$. Besides, $V^{*}\leq V\!\left(X_{*}\right)\leq V\!\left(Z_{*}\right)$, $V_{-}\!\left(Z_{*}^{-}\right)\leq V_{-}\!\left(X_{*}^{-}\right)<+\infty$, so necessarily $V^{
*}=V\!\left(Z_{*}\right)$ must hold, by the definition of $V^{*}$. The proof is complete.\qed
\end{proof}


\section{Multidimensional Diffusion Model With Deterministic Coefficients}\label{ch:Example}

In this section we present an example of an important nontrivial model to which our results can be applied. Let $k$ be a nonnegative integer and let the continuous process
\begin{equation*}
	W=\left\{W_{t}=\left(W^{1}_{t},\ldots,W^{k}_{t}\right)^{\top}\!\textrm{; } 0\leq t \leq T\right\},\footnote{Hereafter, the symbol $\top$ denotes matrix transposition.}
\end{equation*} taking values in $\mathbb{R}^{k}$, be a \mbox{$k$-dimensional} \emph{Brownian motion} (with respect to the probability measure $\mathbb{P}$), which is initialised at zero a.s.. We are assuming the flow of information in the market to be represented by the \emph{natural filtration} of $W$ (which is the augmentation, by all $\mathbb{P}$\hyp{}null sets, of the filtration generated by $W$), $\mathbb{F}^{W}=\left\{\mathscr{F}_{t}^{W}\textrm{; }0\leq t \leq T\right\}$. We further recall that the interest rate is here assumed to be null.

In this particular example, the dynamics of the price process of the $i$\hyp{}th stock $S^{i}=\left\{S^{i}_{t}\text{; } 0 \leq t \leq T\right\}$ is described, under the measure $\mathbb{P}$, by the ItÃ´ process with stochastic differential
\begin{equation}\label{eq:stockprice2}
	dS^{i}_{t}= \mu^{i}\!\left(t\right) S^{i}_{t} dt+\sum_{j=1}^{k} \sigma^{i j}\!\left(t\right) S^{i}_{t} dW^{j}_{t},\qquad S^{i}_{0}=s_{i}>0,
\end{equation}
for any $i \in \left\{1,\ldots,d\right\}$. As made explicit by the notation, here we consider only the case where all the coefficients
\begin{eqnarray*}
	\mu^{i} &=& \left\{\mu^{i}\!\left(t\right)\text{; } 0 \leq t \leq T\right\},\\
	\sigma^{i} &=&\left\{\sigma^{i}\!\left(t\right)=\left(\sigma^{i 1}\!\left(t\right),\ldots,\sigma^{i d}\!\left(t\right)\right)^{\top}\text{; } 0 \leq t \leq T\right\},
\end{eqnarray*}
respectively the \emph{appreciation rate} process and the $\mathbb{R}^{d}$\hyp{}valued \emph{volatility} process of the $i$\hyp{}th risky asset, are deterministic Borelian functions of $t$. Moreover, they satisfy $\int_{0}^{T} \left|\mu^{i}\!\left(t\right)\right| dt+\int_{0}^{T} \sum_{j=1}^{k} \left|\sigma^{i j}\!\left(t\right)\right|^{2} dt < +\infty$, ensuring the existence and uniqueness of strong solutions to the stochastic differential equations (SDE) \eqref{eq:stockprice2}. %
Finally, writing $\sigma\!\left(t\right)$, $0 \leq t \leq T$, to denote the $d\times k$ volatility matrix with entries $\sigma^{i j}\!\left(t\right)$, we assume that $\sigma\!\left(t\right) \sigma\!\left(t\right)^{\top}$ is nonsingular for Lebesgue almost every (Lebesgue a.e.) $t \in \left[0,T\right]$. %

We study two cases separately.

\subsection{Complete Market}

Let us suppose that there are as many risky assets as sources of randomness, that is, $k=d$. Then it is trivial that there exists a uniquely determined $d$-dimensional, deterministic process $\theta=\left\{\theta\!\left(t\right)=\left(\theta^{1}\!\left(t\right),\ldots,\theta^{d}\!\left(t\right)\right)^{\top}\text{; } 0\leq t\leq T\right\}$ such that
\begin{equation*}
	-\mu^{i}\!\left(t\right)=\sum_{j=1}^{d} \sigma^{i j}\!\left(t\right) \theta^{j}\!\left(t\right), \qquad \textrm{for Lebesgue a.e.\ } t \in \left[0,T\right],
\end{equation*}
holds simultaneously for all $i \in \left\{1,\ldots,d\right\}$. In addition, if we assume that the condition $0<\int_{0}^{T} \sum_{i=1}^{d} \left|\theta^{i}\!\left(t\right)\right|^{2} dt<+\infty$ is satisfied, then it is easy to see that the positive local martingale $\left\{\rho_{t}\text{; } 0\leq t\leq T\right\}$ given by
\begin{equation}\label{eq:rhot2}
	\rho_{t}=\exp\!\left\{\sum_{i=1}^{d} \int_{0}^{t} \theta^{i}\!\left(s\right) dW^{i}_{s}-\frac{1}{2}\int_{0}^{t} \sum_{i=1}^{d} \left|\theta^{i}\!\left(s\right)\right|^{2} ds\right\}, \qquad 0 \leq t \leq T, \text{ a.s.,}
\end{equation}
is indeed a martingale under $\mathbb{P}$. Hence, from Cam\-er\mbox{on\hyp{}Mar}\-tin\hyp{}Girsanov theorem and its converse, it is well\hyp{}known that the probability measure $\mathbb{Q}$, with Radon\hyp{}Nikodym derivative $\frac{d\mathbb{Q}}{d\mathbb{P}}=\rho_{T}$ a.s., is the unique e.m.m.\ for $S$, and the process defined by
\begin{equation}\label{eq:BMQ2}
	\widetilde{W}^{i}_{t}=W^{i}_{t}-\int_{0}^{t} \theta^{i}\!\left(s\right) ds, \qquad 0\leq t \leq T,\ i\in \left\{1,\ldots,d\right\},
\end{equation}
is a $d$\hyp{}dimensional Brownian motion on the complete filtered probability space $\left(\Omega,\mathscr{F}_{T}^{W},\mathbb{F}^{W},\mathbb{Q}\right)$. It is also straightforward to check that $\rho_{T}$ is log\hyp{}normally distributed both under $\mathbb{P}$ and under $\mathbb{Q}$. In particular, this implies that both $\rho_{T}$ and $1/\rho_{T}$ have moments of all orders, i.e., $\rho_{T}, 1/\rho_{T} \in \mathscr{W}$, so Assumption~\ref{as:CDFW} holds.

Such a financial market is arbitrage-free and complete, hence any integrable contingent claim is hedgeable. Furthermore, it is trivial to see that there must be some $0<\hat{t}<T$ for which $0<\int_{0}^{\hat{t}} \sum_{i=1}^{d} \left|\theta^{i}\!\left(s\right)\right|^{2} ds<\int_{0}^{T} \sum_{i=1}^{d} \left|\theta^{i}\!\left(s\right)\right|^{2} ds$ holds true, so the vector $\left(\sum_{i=1}^{d} \int_{0}^{\hat{t}} \theta^{i}\!\left(s\right) d W_{s}^{i},\sum_{i=1}^{d} \int_{0}^{T} \theta^{i}\!\left(s\right) d W_{s}^{i}\right)$ has a non-degenerate joint normal distribution. Consequently, we can immediately conclude that the following linear combination of its coordinates
\begin{equation*}
	\frac{\int_{0}^{T} \sum_{i=1}^{d} \left|\theta^{i}\!\left(s\right)\right|^{2} ds}{\int_{0}^{\hat{t}} \sum_{i=1}^{d} \left|\theta^{i}\!\left(s\right)\right|^{2} ds} \sum_{i=1}^{d} \int_{0}^{\hat{t}} \theta^{i}\!\left(s\right) d W_{s}^{i}-\sum_{i=1}^{d} \int_{0}^{T} \theta^{i}\!\left(s\right) d W_{s}^{i}
\end{equation*}
is a $\mathscr{F}^{W}_{T}$-measurable and non-degenerate Gaussian random variable which is independent of $\rho_{T}$. From this, one can easily get a uniform $U_{*}$ independent of $\rho_{T}$, that is, satisfying Assumption~\ref{as:Ustar}.

\subsection{Incomplete Market}

Assume now that $1\leq d<k$, so there exist more sources of risk than traded stocks. It is then clear that a martingale measure for the $d$-dimensional process $S$ is not unique. Indeed, the system of linear equations
\begin{equation}\label{eq:infiniteEMM}
	-\mu^{i}\!\left(t\right)=\sum_{j=1}^{k} \sigma^{i j}\!\left(t\right) \theta^{j}\!\left(t\right), \qquad \textrm{for Lebesgue a.e.\ } t \in \left[0,T\right],\text{ } i\in \left\{1,\ldots,d\right\},
\end{equation}
has infinitely many solutions, and any $\mathbb{R}^{k}$-valued process $\theta=\left\{\theta\!\left(t\right)\textrm{; }0\leq t\leq T\right\}$ satisfying \eqref{eq:infiniteEMM} defines a martingale measure $\mathbb{Q}_{\theta}$ for $S$, 
under suitable conditions. %
Therefore, albeit admitting no arbitrage opportunities, the market is incomplete. 
Nonetheless, it is possible to construct a standard $k$-dimensional Brownian motion (with respect to the probability measure $\mathbb{P}$) starting at zero a.s., which we denote by $\bar{W}=\left\{\bar{W}_{t}=\left(\bar{W}_{t}^{1},\ldots,\bar{W}_{t}^{k}\right)^{\top}\textrm{; }0\leq t \leq T\right\}$, whose natural filtration coincides with that of $W$ and such that the price process $S^{i}$ follows (under $\mathbb{P}$) the dynamics
\begin{equation*}
	dS^{i}_{t}= \mu^{i}\!\left(t\right) S^{i}_{t} dt+\sum_{j=1}^{d} \bar{\sigma}^{i j}\!\left(t\right) S^{i}_{t} d\bar{W}^{j}_{t},
\end{equation*}
for all $i \in \left\{1,\ldots,d\right\}$, where $\bar{\sigma}\!\left(t\right)=\left[\bar{\sigma}^{ij}\!\left(t\right)\right]_{i,j\in \left\{1,\ldots,d\right\}}$ is a deterministic and invertible square matrix of order $d$, for Lebesgue a.e.\ $t \in \left[0,T\right]$, with entries satisfying $\int_{0}^{T} \sum_{i,j=1}^{d} \left|\bar{\sigma}^{ij}\!\left(t\right)\right|^{2}dt<+\infty$. As a consequence, there exists a (unique) deterministic process $\bar{\theta}=\left\{\bar{\theta}\!\left(t\right)=\left(\bar{\theta}^{1}\!\left(t\right),\ldots,\bar{\theta}^{d}\!\left(t\right)\right)^{\top}\textrm{; }0\leq t\leq T\right\}$, taking values in $\mathbb{R}^{d}$, that solves
\begin{equation*}
	-\mu^{i}\!\left(t\right)=\sum_{j=1}^{d} \bar{\sigma}^{i j}\!\left(t\right) \bar{\theta}^{j}\!\left(t\right), \qquad i \in \left\{1,\ldots,d\right\} \textrm{, for Lebesgue a.e.\ } t \in \left[0,T\right].
\end{equation*}
If we impose the additional condition $0<\int_{0}^{T} \sum_{i=1}^{d} \left|\bar{\theta}^{i}\!\left(t\right)\right|^{2}dt<+\infty$, then as above the probability measure $\mathbb{Q}$ given by
\begin{equation*}
	\frac{d\mathbb{Q}}{d\mathbb{P}}=\exp\!\left\{\sum_{i=1}^{d} \int_{0}^{T} \bar{\theta}^{i}\!\left(s\right) d\bar{W}^{i}_{s}-\frac{1}{2}\int_{0}^{T} \sum_{i=1}^{d} \left|\bar{\theta}^{i}\!\left(s\right)\right|^{2} ds\right\}, \text{ a.s.,}
\end{equation*}
is an e.m.m.\ for $S$, and the process $\widetilde{W}=\left\{\widetilde{W}_{t}=\left(\widetilde{W}_{t}^{1},\ldots,\widetilde{W}_{t}^{k}\right)^{\top}\textrm{; }0\leq t \leq T\right\}$, with
\begin{eqnarray*}
	\widetilde{W}_{t}^{i} &=& \bar{W}_{t}^{i}-\int_{0}^{t} \bar{\theta}^{i}\!\left(t\right)ds, \quad \text{ if } i\in\left\{1,\ldots,d\right\},\\
	\widetilde{W}_{t}^{i} &=& \bar{W}_{t}^{i}, \qquad \qquad \qquad \quad \text{ if } i\in\left\{d+1,\ldots,k\right\},
\end{eqnarray*}
is a $k$-dimensional Brownian motion on the probability space $\left(\Omega,\mathscr{F}^{W}_{T},\mathbb{Q}\right)$. Moreover, it is clear that $\mathbb{F}^{W}$ and the natural filtration of $\widetilde{W}$ agree, and that the Radon-Nikodym derivative $\frac{d\mathbb{Q}}{d\mathbb{P}}$ is log-normally distributed (under the measures $\mathbb{P}$ and $\mathbb{Q}$), so again Assumption~\ref{as:CDFW} is verified.

Finally, let us represent by $\mathbb{G}=\left\{\mathscr{G}_{t}\textrm{; }0\leq t \leq T\right\}$ the natural filtration of the $d$-dimensional Brownian motion $\left\{\left(\widetilde{W}^{1}_{t},\ldots,\widetilde{W}^{d}_{t}\right)^{\top}\textrm{; }0\leq t \leq T\right\}$, which is a subfiltration of $\mathbb{F}^{W}$. It is obvious that $\frac{d\mathbb{Q}}{d\mathbb{P}}$ is measurable with respect to $\mathscr{G}_{T}$. Also, by an argument completely analogous to that employed in the complete market case, it is possible to find a $\mathscr{G}_{T}$-measurable, integrable random variable $U_{*}$ so that Assumption~\ref{as:Ustar} is true. In view of the above, it is then easy to show that, in this incomplete financial model, any integrable and $\sigma\!\left(\frac{d\mathbb{Q}}{d\mathbb{P}},U_{*}\right)$-measurable random variable is hedgeable.\\
\\
Hence, taking any hedgeable $B$, the discussion in the preceeding subsections readily leads to the following corollary to Theorem~\ref{Theo:Exist}.
\begin{corollary}
	In the multidimensional diffusion model with deterministic coefficients (\ref{eq:stockprice2}), the behavioural problem is well-posed and there exists an optimal portfolio, provided that $\alpha<\beta$ and $\alpha/\gamma<1<\beta/\delta$.
\end{corollary}

\begin{remark}
	It is possible to construct examples of models with jumps to which Theorem~\ref{Theo:Exist} applies. For instance, if $d=1$ and $S=\left\{S_{t}\text{; } 0\leq t \leq T\right\}$ is a compensated Poisson process, then we can take $\mathbb{Q}=\mathbb{P}$ and Theorem~\ref{Theo:Exist} holds (note that continuity of $F_{\rho}$ is not needed for this existence result). We do not enter into more details here as the examples we can construct look rather artificial.
\end{remark}


\section{Conclusion}\label{ch:Conclusion}

In this work, the optimal portfolio problem was studied for an investor who behaves in accordance with the cumulative prospect theory, assuming a con\-tin\-u\-ous-time market and that asset prices are modelled by general semimartingales.

We addressed, in Section~\ref{ch:WellPosed}, the issue of well-posedness, which is a recurring one within the CPT framework. We focused on the case where the utilities, as well as the probability distortions, are all piecewise-power functions, and we identified several ill-posed cases. We were able to provide necessary and sufficient conditions for well-posedness, which are not only fairly easy to verify, but also have economic interpretations. Moreover, these are very similar to those obtained in the incomplete discrete-time multiperiod model \citep{carassus2011}. One question that remains unanswered, however, is what happens when $\beta=\delta$ or $\alpha=\gamma$.

Next, in Section~\ref{ch:Existence}, we discussed the problem of attainability. Under suitable conditions, we established the existence of an optimal investment strategy, the proof being analogous to that of the multiperiod discrete-time case studied in \citet{carassus2011}. We have not addressed here the question whether this optimal portfolio is unique or not, but there seems to be no hope of uniqueness in this problem. Also, it would be interesting if an optimal solution could be characterised explicitly, in order to compare it with that of EUT.

Finally, in the preceeding section, we concluded this paper with an application of our results to a multidimensional diffusion model with deterministic coefficients. Extending the existence result to diffusion processes with stochastic coefficients is a work in progress.

\appendix

\section{Proofs}

\subsection{Proofs of Section~\ref{ch:WellPosed}}

\subsubsection*{Proof of Lemma~\ref{bd<1aux1}}

Let the function $f:\left[\left.0,+\infty\right)\right. \rightarrow \left[\left.0,+\infty\right)\right.$ be given by $f\!\left(x\right) \colonequals \mathbb{E}\!\left[X\wedge x\right]$. It is clear that $f$ is a non-decreasing function of $x$ and that $\lim_{x \rightarrow +\infty} f\!\left(x\right)=+\infty$. Indeed, to see this let $M>0$ be arbitrary. Since $\mathbb{E}\!\left[X\wedge n\right]\rightarrow \mathbb{E}\!\left[X\right]=+\infty$ by the Monotone Convergence Theorem, there exists some $n_{0} \in \mathbb{N}$ such that $\mathbb{E}\!\left[X\wedge n\right]>M$ for all $n \geq n_{0}$. Given that $f$ is non-decreasing, we conclude $f\!\left(x\right)\geq f\!\left(n_{0}\right)>M$ for all $x\geq n_{0}$, as intended.

Moreover, $f$ is a continuous function on $\left[\left.0,+\infty\right)\right.$. In fact, let $\left\{x_{n}\right\}_{n \in \mathbb{N}}$ be a sequence in $\left[\left.0,+\infty\right)\right.$ converging to some $x\geq 0$. Being convergent, $\left\{x_{n}\right\}_{n \in \mathbb{N}}$ is bounded, that is, there exists some $R>0$ such that $x_{n}\leq R$ for all $n$. Hence, the sequence $\left\{X_{n}\right\}_{n \in \mathbb{N}}$ of random variables $X_{n} \colonequals X\wedge x_{n}$ is dominated by the integrable random variable $X\wedge R$ and converges pointwise to $X\wedge x$ (we recall that the minimum is a continuous function). Therefore, by Lebesgue's Dominated Convergence Theorem, $f\!\left(x_{n}\right)=\mathbb{E}\!\left[X\wedge x_{n}\right] \xrightarrow[n\rightarrow +\infty]{} \mathbb{E}\!\left[X\wedge x\right]=f\!\left(x\right)$ and $f$ is continuous as claimed.

So let us consider $b\geq 0$ arbitrary. It is obvious that $f\!\left(b\right)\leq b$. Besides, since $f\!\left(x\right)\rightarrow +\infty$ as $x\rightarrow +\infty$, there exists some $N>0$ so that $f\!\left(x\right)>b$ for all $x\geq N$. Hence, because $f\!\left(b\right)\leq b < f\!\left(N\right)$ and $f$ is continuous, by the Intermediate Value Theorem (note that $N>b$, otherwise $f\!\left(N\right)\leq f\!\left(b\right)\leq b$ by the monotonicity of $f$) we conclude that $b=f\!\left(a\right)$ for some $a \in \left[b,N\right]$.\qed


\subsubsection*{Proof of Lemma~\ref{EleqL}}
Let $t>0$ be arbitrary (but fixed). Then, for any nonnegative random variable $X$, we have by the monotonicity of the integral that
\begin{eqnarray*}
	\int_{0}^{+\infty} \mathbb{P}\!\left\{X^{b}>y\right\}^{a} dy &=& \int_{0}^{+\infty} \mathbb{P}\!\left\{\left(X^{s}\right)^{\frac{b}{s}}>y\right\}^{a} dy \geq \int_{0}^{t^{\frac{b}{s}}} \mathbb{P}\!\left\{\left(X^{s}\right)^{\frac{b}{s}}>y\right\}^{a} dy\\
	&\geq & \int_{0}^{t^{\frac{b}{s}}} \mathbb{P}\!\left\{\left(X^{s}\right)^{\frac{b}{s}}>t^{\frac{b}{s}}\right\}^{a} dy=t^{\frac{b}{s}} \mathbb{P}\!\left\{\left(X^{s}\right)^{\frac{b}{s}}>t^{\frac{b}{s}}\right\}^{a},
\end{eqnarray*}
where the last inequality follows from the inclusion $\left\{\left(X^{s}\right)^{\frac{b}{s}}>t^{\frac{b}{s}}\right\}\subseteq \left\{\left(X^{s}\right)^{\frac{b}{s}}>y\right\}$ for all $0 \leq y \leq t^{\frac{b}{s}}$. Furthermore, $\mathbb{P}\!\left\{\left(X^{s}\right)^{\frac{b}{s}}>t^{\frac{b}{s}}\right\}=\mathbb{P}\!\left\{X^{s}>t\right\}$, so
\begin{equation*}\label{eq:PleqLt}
	\mathbb{P}\!\left\{X^{s}>t\right\}\leq \frac{1}{t^{\frac{b}{s a}}}\left(\int_{0}^{+\infty} \mathbb{P}\!\left\{X^{b}>y\right\}^{a} dy\right)^{\frac{1}{a}}.
\end{equation*}
Hence,
\begin{eqnarray*}
	\mathbb{E}_{\mathbb{P}}\!\left[X^{s}\right] &=& \int_{0}^{\infty} \mathbb{P}\!\left\{X^{s}>t\right\} dt\leq 1+\int_{1}^{+\infty} \mathbb{P}\!\left\{X^{s}>t\right\} dt\\
	&\leq& 1+\left(\int_{0}^{\infty} \mathbb{P}\!\left\{X^{b}>y\right\}^{a} dy\right)^{\frac{1}{a}}\int_{1}^{+\infty} \frac{1}{t^{\frac{b}{s a}}} dt,
\end{eqnarray*}
and setting the constant $D=\int_{1}^{+\infty} \frac{1}{t^{\frac{b}{s a}}} dt \in \left(0,+\infty\right)$ (recall that $\frac{b}{s a}>1$ by hypothesis), which depends only on the parameters, completes the proof.\qed


\subsubsection*{Proof of Lemma~\ref{Lemeta}}
	We start by noticing that the hypothesis $\alpha<\gamma \leq 1$ implies that $\frac{1}{\gamma}<\frac{1}{\alpha \gamma}$ and $\frac{1}{\gamma}<\frac{1}{\alpha}$. Moreover, since $\alpha<\beta$ and $\delta<\beta$, there exists $\eta$ so that $\max\left\{\alpha,\delta\right\}<\eta<\beta$. In particular, we deduce that $\frac{\eta}{\alpha}>1$, and thus $\frac{\eta}{\alpha \gamma}>\frac{1}{\gamma}$. Hence, we choose $\lambda$ so that $\frac{1}{\gamma}<\lambda<\min\left\{\frac{1}{\alpha},\frac{\eta}{\alpha \gamma}\right\}$. Then, given that $\frac{1}{\lambda \alpha}>1$, there exists some $p$ satisfying $1<p<\frac{1}{\lambda \alpha}$. Finally, we note that $1<\frac{\eta}{\delta}$ and $\frac{\alpha \lambda \gamma}{\delta}<\frac{\eta}{\delta}$ (because $\lambda < \frac{\eta}{\alpha \gamma}$, that is, $\alpha \gamma \lambda < \eta$), so we can take $q$ such that $\max\left\{1,\frac{\alpha \lambda \gamma}{\delta}\right\}<q<\frac{\eta}{\delta}$.
	
	Since for all $y\geq 1$ we have $\mathbb{P}\!\left\{\left(X^{+}\right)^{\alpha}>y\right\}\leq \frac{\mathbb{E}_{\mathbb{P}}\!\left[\left(X^{+}\right)^{\alpha \lambda}\right]}{y^{\lambda}}$ by Chebyshev's inequality, it follows from the monotonicity of the integral that
	\begin{equation*}
		\int_{0}^{+\infty} \mathbb{P}\!\left\{\left(X^{+}\right)^{\alpha}>y\right\}^{\gamma} dy\leq 1+C_{1}\, \mathbb{E}_{\mathbb{P}}\!\left[\left(X^{+}\right)^{\alpha \lambda}\right]^{\gamma},
	\end{equation*}
with the strictly positive constant $C_{1}$ being given by $C_{1}\colonequals \int_{1}^{+\infty} 1/y^{\lambda \gamma} dy$ (we recall that $\lambda \gamma>1$). Also, applying 
H\"older's inequality yields
	\begin{equation*}
		\mathbb{E}_{\mathbb{P}}\!\left[\left(X^{+}\right)^{\alpha \lambda}\right]=\mathbb{E}_{\mathbb{P}}\!\left[\frac{1}{\rho^{1/p}} \rho^{1/p} \left(X^{+}\right)^{\alpha \lambda}\right]\leq C_{2}\,\mathbb{E}_{\mathbb{P}}\!\left[\rho \left(X^{+}\right)^{\alpha \lambda p}\right]^{\frac{1}{p}}=C_{2}\,\mathbb{E}_{\mathbb{Q}}\!\left[\left(X^{+}\right)^{\alpha \lambda p}\right]^{\frac{1}{p}},
	\end{equation*}
	where the constant $C_{2}\colonequals \mathbb{E}_{\mathbb{P}}\!\left[\frac{1}{\rho^{1/\left(p-1\right)}}\right]^{\frac{p-1}{p}}$ is finite and strictly positive. Thus, combining the previous equation and Jensen's inequality for concave functions (we note that $\alpha \lambda p<1$), we obtain
	\begin{eqnarray}
		\mathbb{E}_{\mathbb{P}}\!\left[\left(X^{+}\right)^{\alpha \lambda}\right]^{\gamma} &\leq& C_{3}\,\mathbb{E}_{\mathbb{Q}}\!\left[\left(X^{+}\right)^{\alpha \lambda p}\right]^{\frac{\gamma}{p}}\leq C_{3}\,\mathbb{E}_{\mathbb{Q}}\!\left[X^{+}\right]^{\alpha \lambda \gamma}\nonumber\\
		&=& C_{3}\left(x_{0}+\mathbb{E}_{\mathbb{Q}}\!\left[X^{-}\right]\right)^{\alpha \lambda \gamma} \leq C_{4}+C_{3}\,\mathbb{E}_{\mathbb{Q}}\!\left[X^{-}\right]^{\alpha \lambda \gamma},\qquad \ \ \label{suf:eqaux1}
	\end{eqnarray}
	where the last inequality follows from the trivial inequality $\left|x+y\right|^{a}\leq \left|x\right|^{a}+\left|y\right|^{a}$ for $0<a\leq 1$ (notice that $\lambda<\frac{1}{\alpha}\leq \frac{1}{\alpha \gamma}$ implies $\alpha \lambda \gamma<1$), and $C_{3},C_{4} \in \left(0,+\infty\right)$. Now, we again use H\"older's inequality to see that $\mathbb{E}_{\mathbb{Q}}\!\left[X^{-}\right]^{\alpha \lambda \gamma}\leq C_{5}\,\mathbb{E}_{\mathbb{P}}\!\left[\left(X^{-}\right)^{q}\right]^{\frac{\alpha \lambda \gamma}{q}}$ (here $C_{5}\colonequals \mathbb{E}_{\mathbb{P}}\!\left[\rho^{q/\left(q-1\right)}\right]^{\alpha \lambda \gamma\left(q-1\right)/q}$ is also strictly positive and finite). Moreover, since $\frac{\alpha \lambda \gamma}{q}<\delta<1$, we have by the trivial inequality mentioned above that $\mathbb{E}_{\mathbb{P}}\!\left[\left(X^{-}\right)^{q}\right]^{\frac{\alpha \lambda \gamma}{q}}\leq 2+\mathbb{E}_{\mathbb{P}}\!\left[\left(X^{-}\right)^{q}\right]^{\delta}$. Therefore, these inequalities combined 
with~\eqref{suf:eqaux1} yield
	\begin{eqnarray}
		\mathbb{E}_{\mathbb{P}}\!\left[\left(X^{+}\right)^{\alpha \lambda}\right]^{\gamma} &\leq& C_{6}+C_{7}\,\mathbb{E}_{\mathbb{P}}\!\left[\left(X^{-}\right)^{q}\right]^{\delta}\nonumber\\
		&\leq& C_{6}+C_{7}\left[1+D_{1}\left(\int_{0}^{+\infty} \mathbb{P}\!\left\{\left(X^{-}\right)^{\eta}>y\right\}^{\delta} dy\right)^{\frac{1}{\delta}}\right]^{\delta}\nonumber\\
		&\leq& M_{1}+M_{2}\int_{0}^{+\infty} \mathbb{P}\!\left\{\left(X^{-}\right)^{\eta}>y\right\}^{\delta} dy,\label{eq:EalL1}
	\end{eqnarray}
	where to obtain the second inequality we apply Lemma~\ref{EleqL} above (note that $\frac{\eta}{\delta q}>1$), and the last inequality is again due to the trivial inequality we referred to previously in this proof (with $0<\delta<1$). Furthermore, all constants $C_{6},C_{7},M_{1},M_{2}$ belong to $\left(0,+\infty\right)$.
	
	Hence,
	\begin{equation*}
		\int_{0}^{+\infty} \mathbb{P}\!\left\{\left(X^{+}\right)^{\alpha}>y\right\}^{\gamma} dy \leq L_{1}+L_{2}\int_{0}^{+\infty} \mathbb{P}\!\left\{\left(X^{-}\right)^{\eta}>y\right\}^{\delta} dy,
	\end{equation*}
	with $L_{1}$ and $L_{2}$ positive constants that do not depend on the r.v.\ X (only on the parameters), as intended.\qed


\subsubsection*{Proof of Lemma~\ref{L1L2}}
We start by fixing some $\chi$ satisfying $\frac{1}{s}<\chi<\frac{b}{s a}$. Such a $\chi$ exists since $1<\frac{b}{a}$ implies that $\frac{1}{s}<\frac{b}{s a}$. We also note that, because $\chi a<\frac{b}{s}$, we can choose $\xi$ so that $\chi a<\xi<\frac{b}{s}$.

Let $X$ be an arbitrary nonnegative random variable. Given that $\frac{b}{s \xi}>1$, we know from Lemma~\ref{EleqL} that $\mathbb{E}_{\mathbb{P}}\!\left[X^{\xi}\right]\leq 1+D \left(\int_{0}^{+\infty} \mathbb{P}\!\left\{X^{b}>y\right\}^{s} dy\right)^{1/s}$ for some strictly positive finite constant $D$ (not depending on $X$, but only on the parameters). Therefore, recalling that $s\leq 1$, it follows from the trivial inequality $\left|x+y\right|^{s}\leq \left|x\right|^{s}+\left|y\right|^{s}$ that
\begin{equation}\label{eq:L1L2aux1}
	\mathbb{E}_{\mathbb{P}}\!\left[X^{\xi}\right]^{s}\leq 1+C_{1}\int_{0}^{+\infty} \mathbb{P}\!\left\{X^{b}>y\right\}^{s} dy,
\end{equation}
with $C_{1} \in \left(0,+\infty\right)$. Now, by Jensen's inequality for concave functions (note that $\frac{a \chi}{\xi}<1$), we obtain
\begin{equation}\label{eq:L1L2aux2}
	\mathbb{E}_{\mathbb{P}}\!\left[X^{a \chi}\right]=\mathbb{E}_{\mathbb{P}}\!\left[\left(X^{\xi}\right)^{\frac{a \chi}{\xi}}\right]\leq \mathbb{E}_{\mathbb{P}}\!\left[X^{\xi}\right]^{\frac{a \chi}{\xi}}.
\end{equation}
Moreover, using Chebyshev's inequality, we get
\begin{equation}\label{eq:L1L2aux3}
	\int_{0}^{+\infty} \mathbb{P}\!\left\{X^{a}>y\right\}^{s} dy\leq 1+C_{2}\,\mathbb{E}_{\mathbb{P}}\!\left[X^{a \chi}\right]^{s},
\end{equation}
where the strictly positive constant $C_{2}$ is given by $C_{2}\colonequals \int_{1}^{+\infty} \frac{1}{y^{s \chi}} dy$ (note that $s \chi>1$).

Thus, combining the inequalities~\eqref{eq:L1L2aux1}, \eqref{eq:L1L2aux2} and \eqref{eq:L1L2aux3} above yields
\begin{eqnarray*}
	\int_{0}^{+\infty} \mathbb{P}\!\left\{X^{a}>y\right\}^{s} dy &\leq& 1+C_{2}\left(\mathbb{E}_{\mathbb{P}}\!\left[X^{\xi}\right]^{s}\right)^{\frac{a \chi}{\xi}}\\
	&\leq& 1+C_{2}\left(1+C_{1}\int_{0}^{+\infty} \mathbb{P}\!\left\{X^{b}>y\right\}^{s} dy\right)^{\frac{a \chi}{\xi}}\\
	&\leq& R_{1}+R_{2}\left(\int_{0}^{+\infty} \mathbb{P}\!\left\{X^{b}>y\right\}^{s} dy\right)^{\frac{a \chi}{\xi}}
\end{eqnarray*}
where the last inequality is due to the trivial inequality mentioned above (again we recall that $\frac{a \chi}{\xi}<1$), and the positive constants $R_{1}$, $R_{2}$ depend only on the parameters. Setting $\zeta=\frac{a \chi}{\xi}$ completes the proof.\qed


\subsection{Proofs of Section~\ref{ch:Existence}}

\subsubsection*{Proof of Lemma~\ref{supEtau}}
Let $\lambda>0$ be as in the proof of Lemma~\ref{Lemeta}. We begin by showing that
\begin{equation}\label{eq:supEal}
	\sup_{n \in \mathbb{N}} \mathbb{E}_{\mathbb{P}}\!\left[\left(X_{n}^{+}\right)^{\alpha \lambda}\right]<+\infty.
\end{equation}
Assume by contradiction that $\sup_{n \in \mathbb{N}} \mathbb{E}_{\mathbb{P}}\!\left[\left(X_{n}^{+}\right)^{\alpha \lambda}\right]=+\infty$. Then we can take a subsequence of $\left\{\mathbb{E}_{\mathbb{P}}\!\left[\left(X_{n}^{+}\right)^{\alpha \lambda}\right]\right\}_{n \in \mathbb{N}}$ such that $\mathbb{E}_{\mathbb{P}}\!\left[\left(X_{n_{l}}^{+}\right)^{\alpha \lambda}\right] \rightarrow +\infty$ as $l \rightarrow +\infty$. By equation~\eqref{eq:EalL1} in the proof of Lemma~\ref{Lemeta}, we conclude that $\int_{0}^{+\infty} \mathbb{P}\!\left\{\left(X_{n_{l}}^{-}\right)^{\eta}>y\right\}^{\delta} dy \xrightarrow[l\rightarrow +\infty]{} +\infty$, where $\eta$ is as defined in the proof. Therefore, using Lemma~\ref{L1L2} we also obtain that $V_{-}\!\left(X_{n_{l}}^{-}\right)=\int_{0}^{+\infty} \mathbb{P}\!\left\{\left(X_{n_{l}}^{-}\right)^{\beta}>y\right\}^{\delta} dy \xrightarrow[l\rightarrow +\infty]{} +\infty$, and hence
\begin{eqnarray*}
	V\left(X_{n_{l}}\right) &=& V_{+}\!\left(X_{n_{l}}^{+}\right)-V_{-}\!\left(X_{n_{l}}^{-}\right) \leq L_{1}+L_{2}\int_{0}^{+\infty} \mathbb{P}\!\left\{\left(X_{n_{l}}^{-}\right)^{\eta}>y\right\}^{\delta} dy-V_{-}\!\left(X_{n_{l}}^{-}\right)\\
	&\leq& C_{1}+C_{2} \left(V_{-}\!\left(X_{n_{l}}^{-}\right)\right)^{\zeta}-V_{-}\!\left(X_{n_{l}}^{-}\right) \xrightarrow[l\rightarrow +\infty]{} -\infty,
\end{eqnarray*}
where the first and second inequalities follow, respectively, from Lemma~\ref{Lemeta} and Lemma~\ref{L1L2} ($C_{1}$ and $C_{2}$ are strictly positive constants depending only on the parameters.), and $0<\zeta<1$. But this is absurd, because $\lim_{n \in \mathbb{N}} V\!\left(X_{n}\right)=V^{*}>-\infty$ and therefore any subsequence of $V\!\left(X_{n}\right)$ must also converge to $V^{*}$.

Now we show that \eqref{eq:supEal} implies that $\sup_{n \in \mathbb{N}} V_{-}\!\left(X_{n}^{-}\right)<+\infty$. Indeed, by Chebyshev's inequality, for some $C_{3}\in \left(0,+\infty\right)$, $V_{+}\!\left(X_{n}^{+}\right)\leq 1+C_{3}\,\mathbb{E}_{\mathbb{P}}\!\left[\left(X_{n}^{+}\right)^{\alpha \lambda}\right]^{\gamma}$ for all $n$. Using \eqref{eq:supEal}, we thus get $\sup_{n\in \mathbb{N}} V_{+}\!\left(X_{n}^{+}\right)<+\infty$. Furthermore, since $V\!\left(X_{n}\right) \rightarrow V^{*}$ as $n \rightarrow +\infty$ and any convergent sequence is bounded, we have that $\inf_{n \in \mathbb{N}} V\!\left(X_{n}\right)>-\infty$. Thus,
\begin{equation*}
	\sup_{n \in \mathbb{N}} V_{-}\!\left(X_{n}^{-}\right)\leq \sup_{n \in \mathbb{N}} V_{+}\!\left(X_{n}^{+}\right)-\inf_{n \in \mathbb{N}} V\!\left(X_{n}\right)<+\infty,
\end{equation*}
as intended.

Finally, recalling that $\frac{\beta}{\delta}>1$, we can choose $\xi \in \left(1,\frac{\beta}{\delta}\right)$. Therefore $\frac{\beta}{\delta \xi}>1$, and it follows from Lemma~\ref{EleqL} that there exists some $D\in \left(0,+\infty\right)$ such that
\begin{equation*}
	\mathbb{E}_{\mathbb{P}}\!\left[\left(X_{n}^{-}\right)^{\xi}\right]\leq 1+D\left(V_{-}\!\left(X_{n}^{-}\right)\right)^{\frac{1}{\delta}},
\end{equation*}
for all $n \in \mathbb{N}$, which implies that $\sup_{n \in \mathbb{N}} \mathbb{E}_{\mathbb{P}}\!\left[\left(X_{n}^{-}\right)^{\xi}\right]<+\infty$. In particular, we note that, because $g:\left[\left.0,+\infty\right)\right. \rightarrow \left[\left.0,+\infty\right)\right.$ given by $g\!\left(t\right)\colonequals t^{\xi}$ is a nonnegative, strictly increasing and strictly convex function on $\left[\left.0,+\infty\right)\right.$ satisfying $\lim_{t \rightarrow +\infty} \frac{g\!\left(t\right)}{t}=+\infty$, as well as $\sup_{n \in \mathbb{N}} \mathbb{E}_{\mathbb{P}}\!\left[g\!\left(X_{n}^{-}\right)\right]<+\infty$, by de la Vall\'ee-Poussin theorem %
we conclude that the family $\left\{X_{n}^{-}\right\}_{n \in \mathbb{N}}$ is uniformly integrable.

So we set $\tau=\alpha \lambda \in \left(0,1\right)$. A straightforward application of 
H\"older's inequality (with $p=\frac{\xi}{\tau}>1$) and of the trivial inequality $\left|x+y\right|^{\tau}\leq \left|x\right|^{\tau}+\left|y\right|^{\tau}$ gives
\begin{equation*}
	\mathbb{E}_{\mathbb{P}}\!\left[\left|X_{n}\right|^{\tau}\right]\leq \mathbb{E}_{\mathbb{P}}\!\left[\left(X_{n}^{+}\right)^{\tau}\right]+\mathbb{E}_{\mathbb{P}}\!\left[\left(X_{n}^{-}\right)^{\tau}\right]\leq \mathbb{E}_{\mathbb{P}}\!\left[\left(X_{n}^{+}\right)^{\tau}\right]+\mathbb{E}_{\mathbb{P}}^{\frac{\tau}{\xi}}\!\left[\left(X_{n}^{-}\right)^{\xi}\right],
\end{equation*}
hence $\sup_{n \in \mathbb{N}} \mathbb{E}_{\mathbb{P}}\!\left[\left|X_{n}\right|^{\tau}\right]<+\infty$.\qed


\subsubsection*{Proof of Corollary~\ref{tight1}}
Let $\epsilon>0$ be arbitrary. By Lemma~\ref{supEtau} above, $\sup_{n \in \mathbb{N}} \mathbb{E}_{\mathbb{P}}\!\left[\left|X_{n}\right|^{\tau}\right]=S \in \left[\left.0,+\infty\right)\right.$, for some $\tau \in \left(0,1\right)$. So choosing $M=M\!\left(\epsilon\right)$ such that $M>\left(\frac{S}{\epsilon}\right)^{\frac{1}{\tau}}\geq 0$, and setting $K=\left[-M,M\right]$, by Chebyshev's inequality we obtain
\begin{equation*}
	\mathbb{P}\!\left\{X_{n}\in K^{c}\right\}=\mathbb{P}\!\left\{\left|X_{n}\right|>M\right\}\leq \frac{\mathbb{E}_{\mathbb{P}}\!\left[\left|X_{n}\right|^{\tau}\right]}{M^{\tau}}<\epsilon,
\end{equation*}
which completes the proof.\qed


\begin{acknowledgements}
{AMR is sponsored by the Doctoral Grant SFRH/BD/69360/2010 from the Portuguese Foundation for Science and Technology (FCT). The authors thank an anonymous referee for his/her useful
comments.}
\end{acknowledgements}


%
%

\bibliographystyle{spmpsci}

\end{document}